\documentclass[12pt,showpacs,preprintnumbers,superscriptaddress,nofootinbib]{revtex4-1}
\usepackage{graphicx,amsmath,graphics}
\usepackage{dcolumn}
\usepackage{slashed}
\usepackage{slashbox,booktabs}
\usepackage[usenames]{color}

\usepackage{subcaption}
\usepackage{booktabs}
\usepackage{titlesec}
\usepackage{floatrow}
\usepackage{epstopdf}

\newcommand{\hs}{\hspace*{0.5cm}}

\newcommand{\be}{\begin{equation}}
\newcommand{\ee}{\end{equation}}
\newcommand{\bea}{\begin{eqnarray}}
\newcommand{\eea}{\end{eqnarray}}
\newcommand{\ben}{\begin{enumerate}}
\newcommand{\een}{\end{enumerate}}
\newcommand{\bde}{\begin{widetext}}
\newcommand{\ede}{\end{widetext}}
\newcommand{\nn}{\nonumber}
\newcommand{\crn}{\nonumber \\}

\newcommand{\al}{\alpha}
\newcommand{\la}{\lambda}
\newcommand{\bet}{\beta}

\newcommand{\om}{\omega}
\newcommand{\pa}{\partial}
\newcommand{\+}{\dagger}
\newcommand{\fr}{\frac}
\newcommand{\bc}{\begin{center}}
\newcommand{\ec}{\end{center}}

\newcommand{\de}{\delta}
\newcommand{\De}{\Delta}

\newcommand{\La}{\Lambda}

\newcommand{\gev}{\text{GeV }}	
\begin{document}
\title{ \boldmath
Multiperiod structure of electroweak phase transition in the 3-3-1-1 model}
\author{V. Q. Phong\footnote{vqphong@hcmus.edu.vn}}
\affiliation{  Department of Theoretical Physics, VNUHCM-University of Science, Ho Chi Minh City 700000, Vietnam}
\author{N. T. Tuong\footnote{nttuong.phys@gmail.com}}
\affiliation{  Department of Theoretical Physics, VNUHCM-University of Science, Ho Chi Minh City 700000, Vietnam}
\author{N. C. Thao}
\email{ncthao@grad.iop.vast.ac.vn}
\affiliation{ Graduate University of Science and Technology, Vietnam Academy of Science and Technology, 18 Hoang Quoc Viet, Cau Giay, Hanoi 100000, Vietnam}
\affiliation{Institute of Physics, Vietnam  Academy of Science and Technology, 10 Dao Tan, Ba Dinh, Hanoi 100000, Vietnam}	
\author{H. N. Long}
\email{hoangngoclong@tdtu.edu.vn}
\affiliation{Theoretical Particle Physics and Cosmology Research Group, Advanced Institute for Materials Science, Ton Duc Thang University, Ho Chi Minh City 700000, Vietnam}
\affiliation{Faculty of Applied Sciences, Ton Duc Thang University, Ho Chi Minh City 700000, Vietnam}
\date{\today}

\begin{abstract}
The electroweak phase transition (EWPT) is considered in the framework of 3-3-1-1 model for Dark Matter.
The phase structure within three or two periods is approximated for the theory with many vacuum expectation values (VEVs) at TeV and Electroweak scales. In the mentioned model, there are two pictures.
The first picture containing  two periods of  EWPT,  has a transition $SU(3) \rightarrow  SU(2)$ at 6 TeV scale and another is $SU(2) \rightarrow U(1)$ transition which is the like-standard model EWPT. The second picture is an EWPT structure containing  three periods, in  which  two first periods are similar to those of the first picture and another one is the symmetry breaking process of $U(1)_N$ subgroup.
 Our study leads to the conclusion that EWPTs are the first order phase transitions when new bosons are triggers and their masses
  are within range of some TeVs. Especially, in two pictures, the maximum strength of the $SU(2) \rightarrow U(1)$ phase transition is equal to 2.12 so this EWPT is not strong. Moreover, neutral fermions, which are candidates for Dark Matter and obey the Fermi-Dirac distribution, can be a negative trigger for EWPT.
      However,  they do not make lose the first-order EWPT at TeV scale. Furthermore, in order to be the strong first-order EWPT at TeV scale, the symmetry breaking processes must produce more bosons than fermions or the mass of bosons must be much larger than that of fermions.

\end{abstract}

\maketitle


\section{Introduction}
\label{secInt}

The EWPT is another view of spontaneous symmetry breaking  in Theoretical Particle Physics.
The latter  is a transition of the Higgs field with vanishing VEV to  non-zero one.
The EWPT plays an important role at early stage of expanding universe; and
its  issue is also related to hot topics such as Dark Matter (DM) or Dark Energy. From a micro viewpoint and within the current limits, candidate for  DM  may be a heavy particle. If we accept the symmetry-breaking mechanism as an universal mechanism, then  mass of the DM candidate must also be generated through a phase transition process. Moreover, if the mass of the DM candidate is very large so the phase transition process must take place before the EWPT of the Standard Model (SM) and must also follow the gradually decreasing temperature structure of the universe.

 As in the SM, the EWPT process has only one phase at the energy level
  around 200 GeV. This process is accompanied by mass generation of particles. However, at present, the existence of heavy
   particles is possible only at energy scale larger than 200 GeV. Therefore, the production of these heavy particles
  interacting  with the SM ones, must also be considered.

At present,  the mechanism of symmetry-breaking  is believed to be accurate, but the Higgs potential is not exactly determined because its form is model dependent.

The EWPT consists of an important question of phase transition which must be a strongly first-order phase one. This is the third
Sakharov condition being deviation from thermal equilibrium  Ref.~\cite{sakharov}.
The mentioned condition
together with  B, C, CP violations leads to
solution of the Baryon Asymmetry of Universe (BAU). The
B, C and CP violations can be seen  throughout the sphaleron rate and the Cabibbo-Kobayshi-Maskawa (CKM)-matrix in models Ref.~\cite{SME} or other CP violation sources as neutrino mixing or heavy extra neutrinos via adding see-saw mechanism Ref.~\cite{Fonseca:2016tbn}.

At present, the EWPT is considered at a one-loop level,
  particularly,  in  beyond the  Standard Model.   A new trend nowadays is multi-phase calculations in multi-Higgs scalar potential.

In order to consider the EWPT, we must build the high-temperature effective potential which is usually in the following form

\be V_{eff}=D.(T^2-T^2_0){v}^2-E.Tv^3+\frac{\la_T}{4}v^4,\label{eqintroduc}\ee
where $v$ is the VEV of Higgs  boson. The first order EWPT binds that the strength of phase transition should be
 larger than the unit ($S=\frac{v_c}{T_c}\ge 1$, where $v_c$ is VEV of Higgs field at a critical temperature $T_c$).

The effective potential $V_{eff}$ in Eq. (\ref{eqintroduc}) is a function of temperature and VEVs. It can have one or two minimums
 when the temperature goes down. At $T_c$, the two minimums are separated by a potential barrier, the VEV of Higgs field crosses
  over from  vanishing  VEV to a non-zero VEV. This transition is called the first order phase transition and it can cause large deviations
   from thermal equilibrium.

The EWPT has been calculated in the SM Ref.~\cite{SME} and in some extended models Refs.~\cite{BSM, SMS, dssm, munusm, majorana, thdm, ESMCO, lr,phonglongvan,phonglongvanE,chiang,singlet,mssm1,twostep,1101.4665,Phong:2015vlk}. 
It is  reminded  that DM, heavy particles and neutrino oscillations can be triggers
 of the EWPT Ref.~\cite{elptdm}. The most studies of the EWPT are 
  performed in the framework of the Landau gauge. However gauge also made contributions in EWPT as
 done in Ref. \cite{1101.4665}. It is reminded that in some extended models, Higgs sector consists multi-vacuum structure of which the classical example is  the Two Higgs Doublet model and new models with $SU(5)$
 and $SU(6)$ groups Ref.~\cite{Deppisch:2016jzl} or the $SU(3)_C\otimes SU(2)_L\otimes U(1)_Y (\otimes U(1)_X)$ model as a symmetry of division algebraic ladder operators Ref.~\cite{Furey:2018drh}. This additional Higgs structure can be a new source to answer the BAU puzzles.

Another example of multi-vacuum structure belongs to the models based on $SU(3)_C\otimes SU(3)_L \otimes U(1)_X$ group Refs.~\cite{ppf,r331}
called 3-3-1 models for short. There exist two main versions of the 3-3-1 models: the minimal Ref.~\cite{ppf} and another
with right-handed neutrinos Ref.~\cite{r331}. To provide an explanation for the observed pattern of SM fermion
masses and mixings, various 3-3-1 models with flavor symmetries and radiative seesaw mechanisms have been proposed
in the literature \footnote{With the help of discrete $Z_N$ symmetries, the 3-3-1 model with $\beta = \fr{1}{\sqrt{3}}$ can provide
solutions of neutrino mass and mixing, DM and inflation  Refs.~\cite{n331,plbn}}. However some of them involve non-renormalizable interactions.
 In addition the 3-3-1 models do not give completely desired answer on the DM issue.  In  the recently proposed  3-3-1-1 model Ref.~\cite{3311}
  based on  $SU(3)_C \otimes SU(3)_L \otimes U(1)_X \otimes U(1)_N$ group has a good advantage in explaining DM.
  Phenomena of this model such as DM, inflation,  leptogenesis,  neutrino mass, kinetic mixing effect,
and $B-L$ asymmetry, have been studied in Refs. \cite{3311H,3311HT,3311Dsi,3311DH,3311new}.
The 3-3-1-1 model has three Higgs triplets to generate masses of fermions and the mass of new heavy particle with
 masses around some TeVs.  This model fits  with  candidates for DM. The presence of the above mentioned particles
 might also lead to interesting consequences such as the baryon asymmetry or EWPT which is a subject of this  study.

This article is organized as follows. In section \ref{sec2}, the matter fields and Higgs bosons in the 3-3-1-1 model are briefly reviewed.
 In section \ref{sec3},  the effective potential  having  the contribution from heavy bosons
  a function of temperature and VEVs is derived.
  In section \ref{sec4}, we analysis in details structure of phase transition, find the first order phase transition and show
   constraints on mass of charged Higgs boson in the case without neutral fermions. In section \ref{sec5}, we discuss the role
    of neutral fermions in the EWPT problem. Finally, we summarize and make outlooks in section \ref{sec6}.

\section{Brief review of the 3-3-1-1 model}
\label{sec2}

It is well-known that the SM must be extended and most versions  of the BSM contain heavy particles.
  Within the latter existence, unexplained problems can be caused.
The heavy particles may be a candidate for DM, or  just new ones. The 3-3-1-1 model has many new particles
  inserting  in the multiplets of the gauge group
$ SU(3)_C$ $\otimes SU(3)_L$  $\otimes U(1)_X$ $\otimes U(1)_N$,
where
the latter  is the subgroup   associated with
 the conservation of $B-L$ number
 ~\cite{3311,3311DH,3311Dsi,3311HT,3311H}.

To keep the model being  anomaly free,  the fermion content has to have an equal number of  the $SU(3)_L$ triplets and anti-triplets as follows \cite{3311}
\bea \psi_{aL} &=& \left(
	\nu_{aL},  e_{aL},  (N_{aR})^c
\right)^T \sim \left(1,3, -\fr 1 3,-\fr 2 3\right),\, e_{aR}\sim (1,1, -1,-1),\, \nu_{aR} \sim (1,1,0,-1),
\label{lep}\\
Q_{\al L}&=&\left(d_{\al L}, -u_{\al L},D_{\al L}\right)^T
\sim (3,3^*,0,0),\hs
 Q_{3L} =  \left(
 u_{3L},   d_{3L},  U_L
 \right)^T\sim
\left(3,3,1/3,2/3\right),\crn
 u_{a R}&\sim&\left(3,1,\fr 2 3,\fr 1 3\right),\,  d_{a R} \sim
\left(3,1,-\fr 1 3,\fr 1 3\right),\,
 U_{R}\sim \left(3,1,\fr 2 3,\fr 4 3\right),\, D_{\al R}
\sim \left(3,1,-\fr 1 3,-\fr 2 3\right),\nn \eea
where
 $a = 1, 2, 3$ and $\al=1,2$ are family indices. $N_{aR}$ is neutral fermions playing a role of candidates for DM.
 In (\ref{lep}), the numbers in bracket associated with multiplet correspond to
 number  of members in the  $SU(3)_C$,  $SU(3)_L$ assignment,  its $X$ and $N$ charges, respectively.

The Higgs sector of the model under consideration contains three scalar triplets and one singlet as follows
\bea
\eta &=&  \left(
\eta^0_1\, ,
\eta^-_2\, ,
\eta^0_3
\right)^T\sim (1,3,-1/3,1/3) \, , \hs
 \chi = \left(%\
\chi^0_1\, ,
\chi^-_2\, ,
\chi^0_3%
\right)^T\sim (1,3,-1/3,-2/3)\,  ,\label{chi1}\\
\rho &=& \left(%
\rho^+_1\, ,
\rho^0_2\, ,
\rho^+_3
\right)^T\sim (1,3,2/3,1/3)\, ,\hs
 \phi \sim (1,1,0,2) . \label{phi1}
\eea

 Note that in (\ref{lep}), the lepton and anti-lepton lie in the same triplet. Hence,  lepton number is not conserved and it should be
 replaced with new conserved one $\mathcal{L}$ \cite{cl}. Assuming the bottom element in lepton triplet ($N_{aR}$)
 without lepton number, ones have \cite{3311}
 \be
 B - L = -\fr{2}{\sqrt{3}} T_8 +N\, .
 \label{btl}
 \ee
 Note that in this model, not only leptons but also some scalar fields carry  lepton number as seen in Table \ref{leptonN}

 \begin{table}[htp]
\bc
\caption{
Non-zero lepton number $L$ of fields in the 3-3-1-1 model.}
\begin{tabular}{|c|cccccccccc|}
\hline
Particle & $\nu$ & $e$  & $N$ & $U$ & $D$ & $\eta_{3}$ & $\rho_{3}$ & $\chi_{1}$&$\chi_2$&   $\phi$\\ \hline
$L$ & 1 & 1 & 0& $-1$& 1& $-1$& $-1$ & 1&1&  $-2$ \\
\hline
\end{tabular}
\label{leptonN}
\ec
\end{table}
From  Table \ref{leptonN}, we see that elements at the bottom of $\eta$ and $\rho$ triplets carry lepton number $-1$, while the elements
standing in two first rows of $\chi$ triplet have the opposite one $+1$.

To generate masses for fermions, it is enough that only neutral scalars without lepton number develop  VEV as follows
\bea
\langle \eta \rangle&=&  \left(
\fr{u}{\sqrt{2}}\, ,
0\, ,
\right)^T \, , \hs
\langle \chi \rangle = \left(
0\, ,
0\, ,
\fr{\om}{\sqrt{2}}
\right)^T\, ,\hs
\langle \rho \rangle = \left(
0\, ,
\fr{v}{\sqrt{2}}\, ,
0
\right)^T .
 \label{phi10}
\eea
For the future presentation, let us remind that in the model under consideration,  the covariant derivative is defined as
\be
	D_\mu =\pa_\mu- ig_s t_iG_{i\mu}- igT_i A_{i\mu}- ig_X X B_\mu- ig_N NC_\mu  \, ,\label{2.3}
\ee
  where $G_{i\mu\nu},A_{i\mu\nu},B_{\mu\nu}, C_{\mu\nu}$ and $g_s, g, g_X, g_N$  correspond gauge fields and
   couplings  of  $SU(3)_C,$ $SU(3)_L,$ $U(1)_X $ and $U(1)_N$ groups, respectively.

The Yukawa couplings are  given as
\bea
\mathcal L_{Yukawa}&=&h^e_{ab}\bar{\psi}_{aL}\rho e_{bR}
		+h^\nu_{ab}\bar{\psi}_{aL}\eta \nu_{bR}
		+h'^\nu_{ab}\bar{\nu}^c_{aR}\nu_{bR}\phi
		+h^U\bar Q_{3L}\chi U_R
		+h^D_{\al\beta}\bar{Q}_{\al L}\chi^* D_{\beta R}\crn&&
		+h^u_a\bar Q_{3L}\eta u_{aR}
		+h^d_a\bar Q_{3L}\rho d_{aR}
		+h^d_{ab}\bar Q_{aL}\eta^* d_{bR}
		+h^u_{ab}\bar Q_{aL}\rho^* u_{bR}+H.c..\label{yuk}
\eea
From Eq. (\ref{yuk}), it follows that  masses of the top  and  bottom quarks as follows
\[
m_t=\frac{h_t u}{\sqrt{2}},\
		m_b=\frac{h_b v}{\sqrt{2}},
\]
while masses  of  the exotic  quarks are determined as
\[
	m_U=\frac{\om}{\sqrt{2}}h^U\ ;
	\qquad m_{D_1}=\frac{\om}{\sqrt{2}}h^D_{11}\ ;
	\qquad m_{D_2}=\frac{\om}{\sqrt{2}}h^D_{22}.\]

The Higgs fields are expanded around the VEVs as follows
\bea \eta &= &\langle\eta\rangle+\eta' \, , \eta'=
		\left(\dfrac{S_\eta+iA_\eta}{\sqrt 2}, \, \eta^- , \, \dfrac{S'_\eta+iA'_\eta}{\sqrt 2}\right),\crn
		\rho &= &\langle\rho\rangle+\rho'\, , \rho' =
		\left(\rho^+\, , \dfrac{S_\rho+iA_\rho}{\sqrt 2}\, , \rho'^+\right),\crn
		\chi &= &\langle\chi\rangle+\chi' , \, \chi' =
		+\left(\dfrac{S_\chi+iA_\chi}{\sqrt 2}\, ,\chi^-\, ,\dfrac{S'_\chi+iA'_\chi}{\sqrt 2}\right),\crn
		\phi & =& \langle\phi\rangle+\phi'=\dfrac{\La}{\sqrt 2}+\dfrac{S_4+iA_4}{\sqrt 2}.
\label{hexp}\eea
It is mentioned that the values $u$ and $v$ provide masses for all fermions and gauge bosons  in the SM, while $\om$ gives masses
for the extra heavy quarks and gauge bosons. The value $\La$ plays the role for the $U(1)_N$ breaking at high scale;
and in some cases, it is larger than $\om$.

 The scalar potential for Higgs fields is a function of eighteen parameters
\bea\begin{split}
		V(\rho,\eta,\chi,\phi)=&
		\mu_1^2\rho^\+ \rho +\mu_2^2\chi^\+\chi+\mu_3^2\eta^\+\eta
		+\la_1(\rho^\+\rho)^2
		+\la_2(\chi^\+\chi)^2
		+\la_3(\eta^\+\eta)^2\\&
		+\la_4(\rho^\+\rho)(\chi^\+\chi)
		+\la_5(\rho^\+\rho)(\eta^\+\eta)
		+\la_6(\chi^\+\chi)(\eta^\+\eta)\\&
		+\la_7(\rho^\+\chi)(\chi^\+\rho)
		+\la_8(\rho^\+\eta)(\eta^\+\rho)
		+\la_9(\chi^\+\eta)(\eta^\+\chi)
		+f\varepsilon^{mnp}\eta_m\rho_n\chi_p+H.c)\\&
		+\mu^2\phi^\+\phi+\la(\phi^\+\phi)^2+\la_{10}(\phi^\+\phi)(\rho^\+\rho)
		+\la_{11}(\phi^\+\phi)(\chi^\+\chi)
		+\la_{12}(\phi^\+\phi)(\eta^\+\eta).\end{split}
	\label{pot}\eea

When constructing this Higgs potential, triple scalar self-interactions needs to be limited because it forces us to introduce a $f$ parameter ($f$ has
 a mass dimension the same as $\om$) that can like an interrupt factor for these interactions. In addition, $f$ can be replaced by one Higgs
  field or another interaction among three Higgs fields. Thus, the mentioned interaction will become a fourth- or sixth-order coupling.
  We often do not consider high-order interactions (because these high-order interactions may be difficult to renormalization.
  However they may be related to other hypothetical offending processes). Therefore we can ignore $f$ in this article though it may
   have a different role in other problems. For detailed analysis of the Higgs sector in the model under consideration, the reader is referred to Ref. \cite{3311}.

In this  particular model, the mass of scalar boson depends on not only VEVs, $\mu_{1,2,3}$ and $\la_i, i= 1,2,3,\cdots,12$ but also $f$ parameter.
Note that $f$ increases the mass of bosons \cite{3311}.
Returning to our work, in order to limit  the parameter  number, as above mentioned,  we will  ignore $f$ hereafter.

\subsection{ Higgs boson masses}

Substituting  Eq. (\ref{hexp}) into  Eq. (\ref{pot}) yields
\be	V(\rho,\eta,\chi,\phi)=V_0+V_1+\sum_{i=\rho,\eta,\chi}(V_{S_i}+V_{A_i}+V_i)+V_{S_4}+V_{A_4}+\text{Interaction terms} ,
\label{ve}\ee
where $V_0$ and $V_1$  are the minimum interaction term being independent of  scalar fields and  linear dependent on fields, respectively:
\bea V_0&=&\frac{\la ^2 \La ^4}{4}+\frac{1}{4} \la _{11} \La ^2
		\om ^2+\frac{\la _2 \om ^4}{4}+\frac{\La ^2 \mu
			^2}{2}+\frac{1}{2} \mu _2^2 \om ^2+\frac{\la _3
			u^4}{4}+\frac{1}{4} \la _{12} \La ^2 u^2+\frac{1}{4}
		\la _6 u^2 \om ^2\crn
&+&\frac{1}{2} \mu _3^2 u^2+\frac{1}{4}
		\la _5 u^2 v^2+\frac{\la _1 v^4}{4}+\frac{1}{4} \la
		_{10} \La ^2 v^2+\frac{1}{4} \la _4 v^2 \om
		^2+\frac{1}{2} \mu _1^2 v^2\label{a7},\\
		V_1&=&S_\eta\left[u\mu^2_3+\la_3 u^3+\dfrac{1}{2}\la_5 u v^2+\dfrac{1}{2}\la_6u\om^2+\dfrac{1}{2}\la_{12} u \La^2 \right]
		\crn &+&
		S_\rho\left[v\mu^2_1+\la_1 v^3+\dfrac{1}{2}\la_4 v \om^2+\dfrac{1}{2}\la_5u^2v+\dfrac{1}{2}\la_{10} v \La^2 \right]
		\crn &+&
		S_\chi\left[\om\mu^2_2+\la_2 \om^3+\dfrac{1}{2}\la_4 \om v^2+\dfrac{1}{2}\la_6u^2\om
+\dfrac{1}{2}\la_{11} \om \La^2 \right]
		\crn &+&
		S_4\left[\La\mu^2+\la\La^3+\dfrac{1}{2}\la_{10}v^2\La+\dfrac{1}{2}\la_{11}\La\om^2+
\dfrac{1}{2}\la_{12}  \La u^2 \right].
\eea
Hence, the potential minimization conditions are obtained by
\bea u(\la _{12} \La ^2+\la _6
	\om ^2+2 \mu _3^2+2 \la _3
	u^2+\la _5 v^2)&=&0\label{a9},\\
	\om(\la _{11} \La ^2+2 \la _2
	\om ^2+2 \mu _2^2+\la _6
	u^2+\la _4 v^2)&=&0\label{a10},\\
	v(\la _{10} \La ^2+\la _4
	\om ^2+2 \mu _1^2+\la _5 u^2+2
	\la _1 v^2)&=&0\label{a11},\\
	\La(2 \la \La ^2+\la _{11}
	\om ^2+2 \mu ^2+\la _{12}
	u^2+\la _{10} v^2)&=&0\label{a12}.
\eea
From (\ref{ve}) we get the part for  charged Higgs bosons
\bea V_{\eta}&=&\la_3(\eta^+\eta^-)^2+\left(\dfrac{\La^2\la_{12}}{2}+\dfrac{\la_6\om^2}{2}
+\mu_3^2+\la_3 u^2+\dfrac{\la_5 v^2}{2}+\dfrac{\la_8 v^2}{2}\right)\eta^+\eta^-+\dfrac{1}{2} \la _8 u v \eta
			^{\+ } \rho ^{-}\crn 			
			 &=&\la_3(\eta^+\eta^-)^2+\left(\dfrac{\la _8 v^2}{2}\right)\eta^+\eta^-+\dfrac{1}{2} \la _8 u v \eta
			^{\+ } \rho ^{-},%(\text{theo phương trình (\ref{a9})}).
			\crn
			V_{\chi}& =& \la_2(\chi^+\chi^-)^2+ \left(\dfrac{\La^2\la_{11}}{2}+\la_2\om^2+\mu_2^2
+\dfrac{\la_6 u^2}{2}+\dfrac{\la_7 v^2}{2}+\dfrac{\la_4 v^2}{2}\right)\chi^+\chi^-+\dfrac{1}{2} \la _7 v \om  \chi
			^{-} \rho '^{\+ }\crn 		
			& =& \la_2(\chi^+\chi^-)^2+ \left(\dfrac{\la _7 v^2}{2}\right)\chi^+\chi^-+\dfrac{1}{2} \la _7 v \om  \chi
			^{-} \rho '^{\+ }, \\%(\text{theo phương trình (\ref{a10})}).\\&			
				V_{\rho}& = & \la_1(\rho^+\rho^-)^2+ \left(\dfrac{\La^2\la_{10}}{2}+\dfrac{\la_4\om^2}{2}+\mu_1^2
+\dfrac{\la_5 u^2}{2}+\dfrac{\la_8 u^2}{2}+\la_1 v^2\right)\rho^+\rho^-+\dfrac{1}{2} \la _8 u v \eta ^{-}
				\rho ^\+ 		
				\crn
& =& \la_1(\rho^+\rho^-)^2+ \left(\dfrac{\la _8 u^2}{2}\right)\rho^+\rho^-+\dfrac{1}{2} \la _8 u v \eta ^{-}
				\rho ^{\+ }, %(\text{theo phương trình (\ref{a11})}).
				\crn 		
				V_{\rho'}& =& \la_1(\rho'^+\rho'^-)^2+ \left(\dfrac{\La^2\la_{10}}{2}+\dfrac{\la_7\om^2}{2}+\dfrac{\la_4\om^2}{2}+\mu_1^2+
\dfrac{\la_5 u^2}{2}+\la_1 v^2\right)\rho'^+\rho'^-+\dfrac{1}{2} \la _7 v \om  \chi
				^{\+ } \rho '^{-}\crn
			&	 = & \la_1(\rho'^+\rho'^-)^2+ \left(\dfrac{\la _7 \om ^2}{2}\right)\rho'^+\rho'^-+\dfrac{1}{2} \la _7 v \om  \chi
				^{\+ } \rho '^{-}.%(\text{theo phương trình (\ref{a11})}).
\nn
\eea
From the above equations, after some manipulations,  the mass terms of charged Higgs bosons are given by	
\bea\begin{split}
				V_{Higgs}^{mass}&=\left(\dfrac{\la _8 v^2}{2}\right)\eta^+\eta^-+\dfrac{1}{2} \la _8 u v \eta
				^{\+ } \rho ^{-}+ \left(\dfrac{\la _8 u^2}{2}\right)\rho^+\rho^-+\dfrac{1}{2} \la _8 u v \eta ^{-}
				\rho ^{\+ }\\&\quad
				+ \left(\dfrac{\la _7 v^2}{2}\right)\chi^+\chi^-+\dfrac{1}{2} \la _7 v \om  \chi
				^{-} \rho '^{\+ }+ \left(\dfrac{\la _7 \om ^2}{2}\right)\rho'^+\rho'^-+\dfrac{1}{2} \la _7 v \om  \chi
				^{\+ } \rho '^{-}
				\\&=\dfrac{u^2+v^2}{2}\la _8\left(\dfrac{v\eta^++u\rho^+}{\sqrt{u^2+v^2}} \right)\left(\dfrac{v\eta^-+u\rho^-}{\sqrt{u^2+v^2}}\right)
				\\&\quad+\dfrac{\om^2+v^2}{2}\la _7\left(\dfrac{v\eta^++\om\rho^+}{\sqrt{\om^2+v^2}} \right)\left(\dfrac{v\eta^-+\om\rho^-}{\sqrt{\om^2+v^2}}\right)
				\\&=\dfrac{u^2+v^2}{2}\la _8H_1^+H_1^-+\dfrac{\om^2+v^2}{2}\la _7H_2^+H_2^-
				\\&=m_{H_1}^2H_1^+H_1^-+m_{H_2}^2H_2^+H_2^-,
			\end{split}\nn \eea
where
\bea\begin{split}
				&H_1^\pm=\dfrac{v\eta^\pm+u\rho^\pm}{\sqrt{u^2+v^2}};\\ \\&
				H_2^\pm=\dfrac{v\eta^\pm+\om\rho^\pm}{\sqrt{\om^2+v^2}};
\end{split}
			\qquad
			\begin{split}
				&m_{H_1}^2=\dfrac{u^2+v^2}{2}\la _8;\\ \\&
				m_{H_2}^2=\dfrac{\om^2+v^2}{2}\la _7.
			\end{split}
\eea
		
Similarly, the part of neutral Higgs bosons is given by:	
\bea\begin{split}			&V_{A_4}=\dfrac{\la}{4}A_4^4+\left(\dfrac{1}{2}\la\La^2
+\dfrac{1}{4}\om^2\la_{11}+\dfrac{\mu^2}{2}
+\dfrac{\la_{12}u^2}{4}+\dfrac{\la_{10}v^2}{4}\right)A_4^2
				\\&
				\qquad=\dfrac{\la}{4}A_4^4, \quad %\text{(theo phương trình (\ref{a12})).}
				\\&
				V_{S_4} =\dfrac{\la}{4}S_4^4+\left(\dfrac{3}{2}\la\La^2+\dfrac{1}{4}\om^2\la_{11}+\dfrac{\mu^2}{2}
+\dfrac{\la_{12}u^2}{4}+\dfrac{\la_{10}v^2}{4}\right) {S}_4^2
				\\&	
				\qquad =\dfrac{\la}{4}S_4^4+\la\La^2 {S}_4^2,\quad 
				\\&
				V_{A_\eta} =\dfrac{\la_3}{4}A_{\eta}^4+\left(\dfrac{\La^2\la_{12}}{4}+\dfrac{\la_6\om^2}{4}+\dfrac{\mu_3^2}{2}+\dfrac{\la_3 u^2}{2}+\dfrac{\la_5 v^2}{4}\right)A_\eta^2
				\\&	
				\qquad =\dfrac{\la_3}{4}A_{\eta}^4 ,\quad 
				\\&
				V_{A'_\eta} =\dfrac{\la_3}{4}{A'_\eta}^4+\left(\dfrac{\La^2\la_{12}}{4}+\dfrac{\la_6\om^2}{4}
+\dfrac{\la_9\om^2}{4}+\dfrac{\mu_3^2}{2}+\dfrac{\la_3 u^2}{2}+\dfrac{\la_5 v^2}{4}\right){A'}_\eta^2
				\\&\qquad =\dfrac{\la_3}{4}{A'_\eta}^4+\dfrac{\la_9\om^2}{4}{A'}_\eta^2, \quad 
\end{split}\nn
\eea
\bea\begin{split}
					&	V_{A_\chi} =\dfrac{\la_2}{4}A_{\chi}^4+\left(\dfrac{\La^2\la_{11}}{4}+\dfrac{\la_2\om^2}{2}+\dfrac{\mu_2^2}{2}
+\dfrac{\la_6 u^2}{4}+\dfrac{\la_9 u^2}{4}+\dfrac{\la_4 v^2}{4}\right)A_\chi^2
					\\&	\qquad =\dfrac{\la_2}{4}A_{\chi}^4+\dfrac{\la_9 u^2}{4}A_\chi^2,\quad 
					\\&V_{A'_\chi} =\dfrac{\la_2}{4}{A'_\chi}^4+\left(\dfrac{\La^2\la_{11}}{4}+\dfrac{\la_2\om^2}{2}+\dfrac{\mu_2^2}{2}+\dfrac{\la_6 u^2}{4}+\dfrac{\la_4 v^2}{4}\right){A'}_\chi^2
					\\&\qquad =\dfrac{\la_2}{4}{A'_\chi}^4,\quad %\text{(theo phương trình (\ref{a10})).}
					\\&	V_{A_\rho} =\dfrac{\la_1}{4}A_{\rho}^4+\left(\dfrac{\La^2\la_{10}}{4}+\dfrac{\la_4\om^2}{4}+\dfrac{\mu_1^2}{2}+\dfrac{\la_5 u^2}{4}+\dfrac{\la_1 v^2}{2}\right)A_\rho^2
					\\&	\qquad
					=\dfrac{\la_1}{4}A_{\rho}^4, \quad %\text{(theo phương trình (\ref{a11})).}				
					\\&\qquad=\dfrac{\la_3}{4}S_{\eta}^4+u\la_3S_\eta^3+\la_3 u^2S_\eta^2,\quad %\text{(theo phương trình (\ref{a9})).}
					\\&V_{S'_\chi} =\dfrac{\la_2}{4}{S'_\chi}^4+\om\la_2{S'_\chi}^3+\left(\dfrac{\La^2\la_{11}}{4}+\dfrac{3\la_2\om^2}{2}
+\dfrac{\mu_2^2}{2}+\dfrac{\la_6 u^2}{4}+\dfrac{\la_4 v^2}{4}\right){S'}_\chi^2
					\\&\qquad =\dfrac{\la_2}{4}{S'_\chi}^4+\om\la_2{S'_\chi}^3+\la_2\om^2{S'}_\chi^2,\quad %\text{(theo phương trình ).}
					\\&	V_{S_\rho} =\dfrac{\la_1}{4}S_{\rho}^4+v\la_1{S_\rho}^3+\left(\dfrac{\La^2\la_{10}}{4}+\dfrac{\la_4\om^2}{4}
+\dfrac{\mu_1^2}{2}+\dfrac{\la_5 u^2}{4}+\dfrac{3}{2}\la_1 v^2\right)S_\rho^2
					\\&\qquad =\dfrac{\la_1}{4}S_{\rho}^4+v\la_1{S_\rho}^3+\la_1 v^2S_\rho^2,\quad %\text{(theo phương trình (\ref{a11})).}	
					\label{a16}
\end{split}\eea				
\bea\begin{split}
					&V_{S'_\eta} =\dfrac{\la_3}{4}{S'_\eta}^4+\left(\dfrac{\La^2\la_{12}}{4}+\dfrac{\la_6\om^2}{4}
+\dfrac{\la_9\om^2}{4}+\dfrac{\mu_3^2}{2}+\dfrac{\la_3 u^2}{2}+\dfrac{\la_5 v^2}{4}\right){S'}_\eta^2+\dfrac{1}{2} \la _9 u\om{S'}_\eta S_\chi
					\\&\qquad =\dfrac{\la_3}{4}{S'_\eta}^4+\dfrac{\la_9\om^2}{4}{S'}_\eta^2
+\dfrac{1}{2} \la _9 u\om{S'}_\eta S_\chi,\quad %\text{(theo phương trình (\ref{a9})).}
					\\&	V_{S_\chi} =\dfrac{\la_2}{4}S_{\chi}^4+\left(\dfrac{\La^2\la_{11}}{4}+\dfrac{\la_2\om^2}{2}+\dfrac{\mu_2^2}{2}+\dfrac{\la_6 u^2}{4}+\dfrac{\la_9 u^2}{4}+\dfrac{\la_4 v^2}{4}\right)S_\chi^2
					\\&	\qquad =\dfrac{\la_2}{4}S_{\chi}^4+\dfrac{\la_9 u^2}{4}S_\chi^2.\quad %\text{(theo phương trình (\ref{a10})).}
\end{split}\eea
			
Combination among $S_\chi$ and ${S'}_\eta$ yields
\bea\begin{split}
					V_m(S_\chi,{S'}_\eta)&=\dfrac{\la_9\om^2}{4}{S'}_\eta^2+\dfrac{1}{2} \la _9 u\om{S'}_\eta S_\chi+\dfrac{\la_9 u^2}{4}S_\chi^2\\&
					=\dfrac{\la_9}{4}\left(\om^2{S'}_\eta^2+2u\om{S'}_\eta S_\chi+u^2S_\chi^2\right)\\
					&=\dfrac{\la_9(u^2+\om^2)}{4}\left(\frac{\om{S'}_\eta}{\sqrt{u^2+\om^2}}+\frac{u S_\chi}{\sqrt{u^2+\om^2}}\right)^2\\
					&=\dfrac{\la_9(u^2+\om^2)}{4}\left(H_3\right)^2
					=\dfrac{1}{2}m_{H_3}^2\left(H_3\right)^2,
\end{split}\eea				
where physical  boson $H_3$ is given by
\bea
\begin{split}						H_3=\frac{\om{S'}_\eta}{\sqrt{u^2+\om^2}}+\frac{uS_\chi}{\sqrt{u^2+\om^2}};
\end{split}\qquad \text{with }
\begin{split}
m_{H_3}^2=\dfrac{\la_9(u^2+\om^2)}{2}\label{421}.
\end{split}
\eea
The mass of neutral Higgs bosons is presented  in Table \ref{table0}

\begin{table}[!ht]
		\begin{tabular}{|l|c|c|c|c|c|c|c|} \hline
			Neutral Higgs boson&${S}_4$ &${A'}_\eta$ &$A_\chi$&$S_\eta$&${S'}_\chi$&$S_\rho$&$H_3$
			\\ \hline
			Squared mass&$2\la\La^2$&$\frac{\la_9\om^2}{2}$  & $\frac{\la_9 u^2}{2}$ &$2\la_3 u^2$&$2\la_2\om^2$&$2\la_1 v^2$&$\frac{\la_9(u^2+\om^2)}{2}$
			\\ \hline
		\end{tabular}
		\caption{The neutral Higgs boson masses.}\label{table0}
\end{table}
Remember that the massless Goldstones bosons are: X $A_4, A_\eta, A'_\chi, A_\rho$ in neutral scalar sector and two massless combinations orthogonal
 to the charged Higgs bosons.
It is noted that at the limit $f\longrightarrow 0$,  the results given in \cite{3311H,3311Dsi,3311H,3311HT}
 are consistent with  those of this study.

\subsection{Gauge boson sector}

The gauge bosons obtain masses when the scalar fields develop the VEVs. Therefore, their mass Lagrangian is given by

\[\mathcal{L}^{\mathrm{gauge}}_{\mathrm{mass}}=\sum_\Phi (D^\mu\langle \Phi \rangle)^\+(D_\mu \langle \Phi \rangle ).\]
Substituting the scalar multiplets $\eta$, $\rho$, $\chi$ and $\phi$ with their covariant derivative, we obtain
\bea \mathcal{L}^{\mathrm{mass}}_{\mathrm{gauge}} &=&  \frac{ g^2u ^2 }{8}\left[ \left( A_{3\mu }
 + \frac{A_{8\mu } }{\sqrt{3}} - \frac{2}{3}t_X B_\mu   + \frac{2}{3}t_N C_\mu   \right)^2  + 2W^+_\mu  W^{-\mu}
  + 2X^{0*}_\mu X^{0\mu}  \right]\crn
&&+\frac{g^2v ^2 }{8}\left[ \left(  - A_{3\mu }  + \frac{A_{8\mu} }{\sqrt{3}} + \frac{4}{3}t_X B_\mu
 + \frac{2}{3}t_N C_\mu \right)^2  + 2W^+_\mu W^{-\mu}  + 2Y^+_\mu Y^{-\mu}   \right]\crn
&&+ \frac{g^2\om ^2 }{8}\left[ \left(  - \frac{2A_{8\mu } }{\sqrt{3}} - \frac{2}{3}t_X B_\mu   -
 \frac{4}{3}t_N C_\mu   \right)^2  + 2Y^+_\mu Y^{-\mu}   + 2X^{0*}_\mu X^{0\mu}  \right]\crn
&& + 2g_N^2  \La ^2 C_\mu ^2,\nn\eea
where we have denoted $t_X\equiv \fr{g_X}{g}$, $t_N\equiv \fr{g_N}{g}$,
and
\bea
W_\mu^{\pm}=\fr{A_{1\mu} \mp i A_{2\mu}}{\sqrt 2},\hs X^{0,0*}_\mu= \fr{A_{4\mu}\mp i A_{5\mu}}{\sqrt 2},\hs
Y_\mu^\mp=\fr{A_{6\mu} \mp i A_{7\mu}}{\sqrt 2}. \label{nonherg}
\eea
The mass Lagrangian  can be rewritten as  \cite{3311DH,3311Dsi,3311H,3311HT}
\bea
\mathcal{L}_{\mathrm{mass}}^{\mathrm{gauge}}
&=& \frac{g^2 }{4}\left(u^2  + v^2\right) W^+ W^-+ \frac{g^2 }{4}  \left(v^2  + \om ^2 \right) Y^ +  Y^ -
 + \frac{g^2 }{4} \left( u^2  + \om^2  \right) X^{0*}X^0 \crn
&&+ \frac{1}{2}\left(A_{3} \  A_{8}  \  B \ C  \right) M^2
\left(\begin{array}{c}
	A_{3 }   \\
	A_{8 }   \\
	B  \\
	C
\end{array} \right),\nn \eea
where the Lorentz indices have been omitted and should be understood. The squared-mass matrix of the neutral gauge bosons is found to be:
\small{
\bea M^2=
\fr{g^2}{2}\left(
\begin{array}{cccc}
	\fr 1 2 (u^2 + v^2) &
	\fr{u^2 - v^2}{2\sqrt 3} &
	-\fr{t_X(u^2 +2 v^2)}{3} &
	\fr{t_N (u^2 - v^2)}{3} \\
	\fr{u^2 - v^2}{2\sqrt 3} &
	\fr 1 6 (u^2 + v^2 + 4 \om^2) &
	-\fr {t_X(u^2 -2(  v^2  +\om^2))}{3\sqrt 3} &
	\fr {t_N(u^2 +v^2 +4\om^2)}{3\sqrt 3} \\
	-\fr{t_X(u^2 +2 v^2)}{3} &
	-\fr {t_X(u^2 -2(  v^2  +\om^2))}{3\sqrt 3} &
	\fr {2}{9}t_X^2 (u^2 +4 v^2  +\om^2) &
	-\fr {2}{9}t_X t_N (u^2 -2( v^2 +\om^2 )) \\
	\fr{t_N (u^2 - v^2)}{3} &
	\fr {t_N (u^2 +v^2 +4\om^2)}{3\sqrt 3} &
	-\fr {2}{9}t_X t_N (u^2 -2( v^2 +\om^2 )) &
	\fr {2}{9}t_N^2 (u^2 + v^2 +4 ( \om^2+9 \La^2))  \\
\end{array}
\right).\nn
\eea
}

The non-Hermitian gauge bosons $W^{\pm}$, $X^{0,0*}$ and $Y^\pm$
 are physical fields with corresponding masses:
\[
m^2_W
=\fr{ g^2}{4} (u^2 + v^2 ),\hs m^2_X=\fr{ g^2}{4} (u^2 + \om^2),\hs
m^2_Y =\fr{ g^2}{4} (v^2 + \om^2) .\]
Because of the constraints $u, v\ll \om$, we have $m_W\ll m_{X}\simeq m_{Y}$. The $W$ boson  is identified as the SM $W$ boson. It follows
\[ u^2+v^2=(246\ \mathrm{GeV})^2.\]
 The $X$ and $Y$ fields are the new gauge bosons with the large masses  given in the $\om$ scale. The physical charged gauge bosons and
  their masses are summarized in Table \ref{table23}

\begin{table}[!ht]
		\begin{tabular}{l|c|c|c} \hline
		Gauge boson &$W$ & $Y$ & $X$
			\\ \hline
			Squared mass&$\frac{g^2}{4}(u^2+v^2)$
			& $\frac{g^2}{4}(\om^2+v^2)$
			& $\frac{g^2}{4}(\om^2+u^2)$
			\\ \hline
		\end{tabular}
		\caption{The mass of charged gauge bosons.}
		\label{table23}
\end{table}

It is worth mentioning that after diagonalization, in the obtained masses of gauge bosons, there is  no mixing among
the    VEVs, i.e., in the expression of squared  masses,  there are no  terms such as $u v, u \om, v \om$, etc. For more details, the reader is referred to  Ref. \cite{phonglongvanE}.

From aforementioned analysis, it follows that the phenomenological aspects of the 3-3-1-1 model can be
 divided into two pictures corresponding to different domain values of VEVs.

\subsubsection{Picture (i):  $\La \sim \om \gg v \sim u$}

The physical neutral gauge bosons are derived through the following transformation
 $(A_3,\ A_8,\ B,\ C)\rightarrow (A,\ Z,\ Z_2,\ Z_1)$:
\bea \left(\begin{array}{c}
	A_3\\
	A_8\\
	B\\
	C
\end{array}\right)=
U_1U_2U_3\left(\begin{array}{c}
	A\\
	Z\\
	Z_2\\
	Z_1
\end{array}\right) .\nn
\eea
The above diagonalization is realized
through three steps \cite{3311DH,3311Dsi,3311H,3311HT},
\bea
&&\text{The first step: } M'^2=U^T_1 M^2 U_1,\crn
&&\text{The second step: } M''^2 = U^T_2 M'^2 U_2,\crn
&&\text{The final step: } M'''^2 = U^T_3 M''^2 U_3 = \mathrm{diag}(0,m^2_{Z},m^2_{Z_2},m^2_{Z_1}) ,
\nn\eea
where
\bea
U_1&=&\left(
\begin{array}{cccc}
	s_W & c_W & 0 & 0\\
	-\fr{s_W}{\sqrt{3}} & \fr{s_W t_W}{\sqrt{3}} & \sqrt{1-\fr{t^2_W}{3}} & 0 \\
	c_W\sqrt{1-\fr{t^2_W}{3}} & -s_W\sqrt{1-\fr{t^2_W}{3}}& \fr{t_W}{\sqrt{3}} & 0\\
	0 & 0 & 0 & 1
\end{array} \right) , \crn
&U_2& \simeq \left(
\begin{array}{ccc}
	1 & 0 & 0  \\
	0 &  1 & \mathcal{E} \\
	0 & -\mathcal{E}^T & 1 \\
\end{array}
\right), U_3=\left(
\begin{array}{cccc}
	1 & 0 & 0& 0\\
	0 & 1& 0 & 0\\
	0 & 0 & c_\xi & -s_\xi \\
	0 & 0& s_\xi & c_\xi
\end{array}
\right) .\label{eqt}\eea
In Eq.(\ref{eqt}), the $\mathcal{E}$ is a two-component vector  given by \cite{3311DH,3311Dsi,3311H,3311HT}
\bea
\mathcal{E}_1 &=& -\fr{\sqrt{4 t_X^2+3} \{3 \La^2 [(2 t_X^2-3) u^2+(4 t_X^2+3) v^2]+t_X^2 \om^2 (u^2+v^2)\}}{4 \La^2 (t_X^2+3)^2 \om^2}\ll 1,\crn
\mathcal{E}_2 &=& \fr{t_X^2 \sqrt{4 t_X^2+3} (u^2+v^2)}{8 \La^2 (t_X^2+3)^{3/2} t_N}\ll 1,\crn
t_{2\xi}&\simeq &\fr{4\sqrt{3 + t_X^2} t_N \om^2}{(3 + t_X^2) \om^2- 4 t_N^2 (\om^2 + 9 \La^2)} ,\crn
s_W&=&\fr{\sqrt{3}t_X}{\sqrt{3+4t^2_X}} \simeq \sqrt{0.231} .\nn
\eea
Finally we obtain the masses of neutral gauge bosons as follows
{\footnotesize\bea
m_{Z}^2 &\simeq& \fr{g^2 (u^2 + v^2)} {4 c_W^2},\label{kluongznhe}\\
m_{Z_1}^2 & \simeq &
\fr{g^2} {18}\left((3 + t_X^2) \om^2+ 4 t_N^2 (\om^2 + 9 \La^2) + \sqrt{((3 + t_X^2) \om^2- 4 t_N^2 (\om^2 + 9 \La^2))^2
	+ 16 (3 + t_X^2)t_N^2 \om^4}\right),\label{kluongz1nhe}\\
m_{Z_2}^2& \simeq&
\fr{g^2} {18}\left((3 + t_X^2) \om^2+ 4 t_N^2 (\om^2 + 9 \La^2) -\sqrt{((3 + t_X^2) \om^2- 4 t_N^2 (\om^2 + 9 \La^2))^2
	+ 16 (3 + t_X^2)t_N^2 \om^4}\right) .\label{kluongz2nhe}
\eea}
From the experimental data
$\De \rho < 0.0007$, ones get $u/\om < 0.0544$ or $\om>3.198$ TeV \cite{3311} (provided that $u=246/\sqrt{2}$ GeV as mentioned). Therefore,
 the value of $\om$ results in the TeV scale as expected.

 It has been showed that the ordinary 3-3-1 models are only effective theory, as the B - L charge and the unitarity are violated~\cite{3311D1}. For the case $\La \gg w $, the limit of the 3-3-1 breaking scale followed from flavor
 changing neutral currents as well as LEPII searches
 is $w > 3.6$ TeV.  Due to extra $U(1)_N$ subgroup, the kinetic terms give an effect on the $\rho$ parameter.
 It is well known that the radiative correction of heavy particles groups into Peskin-Takeuchi $S, T, U$ parameters ~\cite{peskintakeuchi,zstu}.  In the frameworks of the 3-3-1 models, the above parameters were investigated in Refs.~\cite{3311D1,stu331,3311Dsi}
  It was remarked that
 if $\La \gg w$, the $\De \rho$ relating to the oblique parameter $T$ depends only on $w$ and $\bet$
 - the parameter appeared in the electric charge operator, not on
 $\La, \fr{g_N} g $ and on $\de$- the coefficient of  mixing  between $B_{\mu \nu}$ and  $C_{\mu \nu}$~\cite{3311Dsi}. In the case of $\La \gg w$, the result is the same as before, i.e., $w > 3.6$ TeV. In the case $\La = 2 w$, the value of $w$ ranges from 3 to 3.5 TeV.

  From LHC searches, it follows that the lower bound on the $Z^{\prime }$
  boson mass in 3-3-1 models is around $2.5$ TeV \cite{Salazar:2015gxa}.
 Hence, the 3-3-1 scale $\om$ is  about $6.3$ TeV.   In addition, from the decays $B_{s,d}\rightarrow \mu ^{+}\mu ^{-}$
 and $B_{d}\rightarrow K^{\ast }\, (K)\, \mu ^{+}\mu ^{-}$
 \cite{CarcamoHernandez:2005ka,Martinez:2008jj,Buras:2013dea,Buras:2014yna,Buras:2012dp} it  follows that the lower limit on the $Z^{\prime }$ boson mass
 ranges from $1$ TeV  to $3$ TeV.
  Hence, both  ordinary 3-3-1 and 3-3-1-1 models provide the similar bound on $\om$.

\subsubsection{Picture (ii): $\La \gg \om \gg v \sim u$}

If we assume $\La \gg \om \gg u\sim v$, three gauge bosons are derived as  \cite{3311DH,3311Dsi,3311H,3311HT}
\bea
m_{Z}^2 &\simeq &\fr{g^2 (u^2 + v^2)} {4 c_W^2} ,\label{chuan1}\\
m_{Z_1}^2 &\simeq & 4 g^2 t_2^2 \La^2 ,\label{chuan2}\\
m_{Z_2}^2&\simeq& \fr{g^2c_W^2 \om^2}{(3 - 4 s_W^2)} .\label{chuan3}
\eea
From the Table (\ref{table23}) and Eqs.(\ref{chuan1}, \ref{chuan2}, \ref{chuan3}), the $W^\pm$ boson and the $Z$ boson are
 recognized as two  famous gauge bosons in the SM. Now we turn to the main object - the effective potential.

\section{Effective potential}\label{sec3}

Within  the above assumption, the  Higgs potential  is given as follows\cite{3311,3311DH,3311Dsi,3311H,3311HT},
\bea
\begin{split}
		V(\rho,\eta,\chi,\phi)=&
		\mu_1^2\rho^\+ \rho +\mu_2^2\chi^\+\chi+\mu_3^2\eta^\+\eta
		+\la_1(\rho^\+\rho)^2
		+\la_2(\chi^\+\chi)^2
		+\la_3(\eta^\+\eta)^2\\&
		+\la_4(\rho^\+\rho)(\chi^\+\chi)
		+\la_5(\rho^\+\rho)(\eta^\+\eta)
		+\la_6(\chi^\+\chi)(\eta^\+\eta)\\&
		+\la_7(\rho^\+\chi)(\chi^\+\rho)
		+\la_8(\rho^\+\eta)(\eta^\+\rho)
		+\la_9(\chi^\+\eta)(\eta^\+\chi)
		%+f\varepsilon^{mnp}\eta_m\rho_n\chi_p+H.c)
		\\&
		+\mu^2\phi^\+\phi+\la(\phi^\+\phi)^2+\la_{10}(\phi^\+\phi)(\rho^\+\rho)
		+\la_{11}(\phi^\+\phi)(\chi^\+\chi)
		+\la_{12}(\phi^\+\phi)(\eta^\+\eta)
\end{split} ,\label{t1}
\eea
from which, ones obtain $V_0$ depending on VEVs :
\bea\begin{split}
		V_0=&\frac{\la \phi_\La ^4}{4}+\frac{1}{4} \la _{11} \phi_\La ^2
		\phi_\om ^2+\frac{\la _2 \phi_\om ^4}{4}+\frac{\phi_\La ^2 \mu
			^2}{2}+\frac{1}{2} \mu _2^2 \phi_\om ^2+\frac{\la _3
			\phi_u^4}{4}+\frac{1}{4} \la _{12} \phi_\La ^2 \phi_u^2+\frac{1}{4}
		\la _6 \phi_u^2 \phi_\om ^2\\&
		+\frac{1}{2} \mu _3^2 \phi_u^2+\frac{1}{4}
		\la _5 \phi_u^2 \phi_v^2+\frac{\la _1 \phi_v^4}{4}+\frac{1}{4} \la
		_{10} \phi_\La ^2 \phi_v^2+\frac{1}{4} \la _4 \phi_v^2 \phi_\om
		^2+\frac{1}{2} \mu _1^2 \phi_v^2 .
	\end{split} \label{t2}
\eea
Here $V_0$ has quartic form like in the SM, but it depends on four variables $\phi_\La, \phi_\om,
\phi_u$, $\phi_v$, and has the mixing terms between  them. However, developing the potential (\ref{t1}), we obtain four minimum equations. Therefore, we can transform the mixing between four variables to the form depending only on $\phi_\La, \phi_\om, \phi_u$ and  $\phi_v$.  Let us explain this point in details. The minimum conditions eliminate the mixing only inside the actual VEV. The above mentioned mixing is due to couplings between fields of the Higgs potential. The Higgs masses do not have the mixing of VEVs when the fields are inside actual VEV. Outside VEV,  the fields do not have masses. Hence the symmetry is restored; and consequently the EWPT does not exist.

Furthermore, importantly, there are the mixings of VEVs because of the unwanted terms
	such as $\la_4(\rho^\+\rho)(\chi^\+\chi)$, $\la_5(\rho^\+\rho)(\eta^\+\eta)$, $\la_6(\chi^\+\chi)(\eta^\+\eta)$, $\la_7(\rho^\+\chi)(\chi^\+\rho)$, $\la_8(\rho^\+\eta)(\eta^\+\rho)$, $\la_9(\chi^\+\eta)(\eta^\+\chi)$, $\la_{10}(\phi^\+\phi)(\rho^\+\rho)$, $\la_{11}(\phi^\+\phi)(\chi^\+\chi)$ and $\la_{12}(\phi^\+\phi)(\eta^\+\eta)$ in Eq.~(\ref{t1}).  To satisfy the generation of inflation
	with $\phi$-inflaton \cite{3311H,3311DH}, the values $\la_{10,11,12}$ can be small, is about $10^{-10}-10^{-6}$. Thus,  $\la_{4,5,6,7,8,9}$ must be also small to make  the corrections of high order interactions of the Higgs will not be divergent.

\begin{figure}[!ht]	
	\includegraphics[scale=1.2]{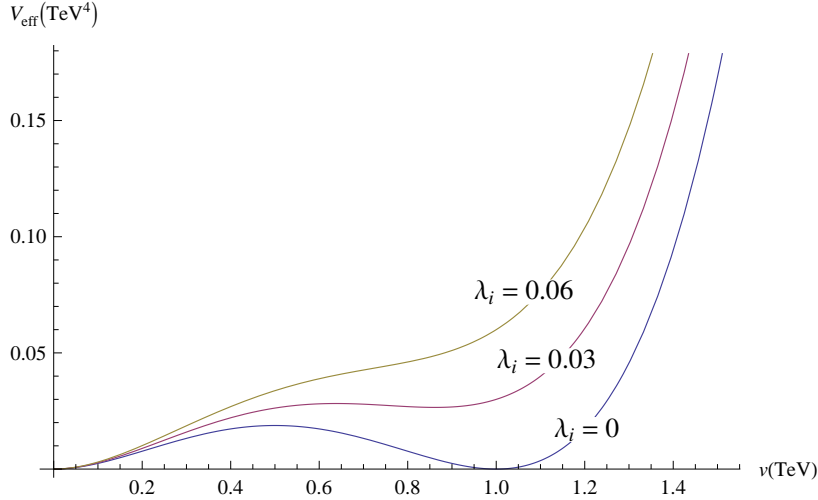}
	\caption{The contours of  the effective potential in (\ref{eqthem})
as a function of $v$ for some values of $\la_i$ as $\la=0.3, D=0.3,E=0.6, \La^2+\om^2+v^2=1\, \textrm{TeV}^2$}\label{mixing}
\end{figure}

In general, if we did not neglect these mixings, $V_0$ will have additional components $\La v$, $\La \om$, $\om v$, $u v$. At the temperature T, for instance, the effective potential depending on VEV $v$, will
be a example form:
\be
V_ {eff} (v) = \la v^4-Ev^3 + Dv^2 + \la_k. \om^2 v^2 + \la_j. \La^2 v^2 +u^2.v^2\approx \la v^4-Ev^3 + Dv^2 + \la_i. (\om^2 + \La^2+u^2) v^2
\label{eqthem}
\ee
The contours of  the effective potential in (\ref{eqthem}) at $\om^2+\La^2+u^2 =1 \, \textrm{TeV}^2$
as a function of $v$ for some values of $\la_i$ is plotted in Fig. \ref{mixing}

 From figure~\ref{mixing}, we see that  at arbitrary temperature $T$
when $\la_i, i = 4, .., 9$ increases, the second minimum of the effective potential fades. For a first order phase transition, the value of $\la_i$ is not too large, so that the potential still has two minima. We observe that if $\la_i$ is enough small to have a second minimum, at arbitrary temperature, the shape of the effective potential remains the same in the absence of $\la_i$. Therefore, we have one more reason to assume that $\la_i$ must be small and  this mixing can be neglected.
Hence, we can write $V_0(\phi_\La,\phi_\om,\phi_u,\phi_v)=V_0(\phi_\La)+V_0(\phi_\om)+V_0(\phi_u)+V_0(\phi_v)$ and ignore the mixing of different VEVs, otherwise our phase transitions will be very complex or distort.

In order to derive effective potential, we need the mass spectrum of fields. Starting from the   Lagrangian of the scalars (both kinetic and potential terms) and Yukawa interactions, and expanding Higgs fields around VEVs, we obtain the mass terms for all fields in the 3-3-1-1 model.

The gauge sector in the  3-3-1-1 has ten gauge bosons: the photon and  nine massive gauge bosons. The latter includes two massive
 like the SM $Z$ and $W^\pm$ bosons,  and two new heavy neutral  $Z_1,Z_2$ bosons, the charged gauge bosons $ Y^\pm$ and
   the neutral non-Hermitian bosons: $X^{0,0^*}$.  The Higgs sector contains  four charged Higgs bosons $H_1^\pm, H_2^\pm$, seven neutral
    Higgs bosons  $S_4, A'_\eta, A_\chi, S_\eta, S'_\chi, S_\rho, H_3$. The model  consists of  four heavy quarks $U, D_1, D_2$, top quark.
     Masses of fields in the 3-3-1-1 model are presented in Table \ref{photong}.

\begin{table}[!ht]
	\caption{Mass formulas of particles in 3-3-1-1 model.}
	\centering
	\scalebox{1}{
		\begin{tabular}{ |l|c|c|c|c|c|c|c|c|}
			\hline
			Boson&$W^\pm$&$Y^\pm$&$X$&$Z$&$Z_1$&$Z_2$
			\\ \hline
			Squared mass & $\frac{g^2}{4}(\phi_u^2+\phi_v^2)$& $ \frac{g^2}{4}(\phi_\om^2+\phi_v^2)$ &
 $\frac{g^2}{4}(\phi_\om^2+\phi_u^2) $ & & & \\
			Picture (i)&&&&Eq.(\ref{kluongznhe}) & Eq.(\ref{kluongz1nhe}) & Eq.(\ref{kluongz2nhe})\\
			Picture (ii)&&&&$\frac{g^2(\phi_u^2+\phi_v^2)}{4c_W^2}$ & $4g^2t_N^2\phi_\La^2$ & $\frac{g^2c^2_W\phi_\om^2}{3-4s_W^2}$\\
			\hline
			
			Neutral Higgs boson&$S'_\chi$&$S_\rho$& $S_4$&$A'_\eta$&$A_\chi$&$S_\eta$ \\ \hline
			 Squared mass&$2\la_2\phi_\om^2$&$2\la_1\phi_v^2$&$2\la\phi_\La^2$&
$\frac{\la_9\phi_\om^2}{2}$&$\frac{\la_9 \phi_u^2}{2}$&$2\la_3 \phi_u^2$\\
			\hline
			
			Charged Higgs boson&$H_1$&$H_2$&$H_3$& & & \\
			Complex Higgs boson&&&Complex Higgs boson&&&\\ \hline
			Squared mass&$\frac{\phi_u^2+\phi_v^2}{2}\la _8$& $\frac{\phi_\om^2+\phi_v^2}{2}\la_7$&
$\frac{\la_9(\phi_u^2+\phi_\om^2)}{2}$
			& & &\\
			\hline
			
			Quark& U&$D_1$&$D_2$&Top&&\\ \hline
	      Squared mass&$\frac{1}{2}{h^U}^2\phi_\om^2$&$\frac{1}{2}h^{D^2}_{11}\phi_\om^2$&
$\frac 1 2 h^{D^2}_{22}\phi_\om^2$&$\frac{1}{2}h^2_t \phi_u^2$&&\\ \hline
		\end{tabular}\label{photong}}
	\end{table}
From the mass spectra, we can split masses of particles into four parts as follows
\be
m^2(\phi_\La,\phi_\om,\phi_u,\phi_v)=m^2(\phi_\La )+m^2( \phi_\om )+m^2( \phi_u )+m^2( \phi_v) .\label{t3}
\ee

Taking into account Eqs.  (\ref{t2}) and (\ref{t3}), we can also split the effective potential into four parts
\[
V_{eff}(\phi_\La,\phi_\om,\phi_u,\phi_v)=V_{eff}(\phi_\La )+
V_{eff}( \phi_\om )+V_{eff}( \phi_u )+V_{eff}( \phi_v) .
\]
It is difficult to study the electroweak phase transition with four VEVs, so we assume $\phi_\La \approx\phi_\om, \phi_u \approx\phi_v$
 over space-times. Then, the effective potential becomes	
\[
V_{eff}(\phi_\La,\phi_\om,\phi_u,\phi_v)=V_{eff}( \phi_\om )+V_{eff}( \phi_u ) .\]

  From Table \ref{photong}, it follows that the squared  masses of  gauge and  Higgs bosons  are split into three separated parts corresponding to three SSB stages.
 It is consistent with the analysis given  in Ref. \cite{phonglongvanE}.

It is interesting to note that the way of  splitting  into two or three phases performed in this paper is   similar to that
in Ref.  \cite{JHEP03(2018)114}.
However, in this paper, the multi-periodicity of phase transition is shown  more transparently   both  in its title and as well as in  content.

\section{Electroweak phase transition without neutral fermion}\label{sec4}

Taking phase transitions in this model into account,  it is important to find the activity domain of $\om$, $\La$, $u$ and $ v$.
Looking at  data in Ref. \cite{limit,pdg}, we arrive to assumption:   $m_{Z_2}\ge 2.2$ TeV. In addition, from
 Ref. \cite{3311}, we also assume $m_{Z_2}<2.5$ TeV. Hence
\be
 2.2 \, \text{TeV } \le m_{Z_2}\le 2.5 \textrm{ TeV} .\label{limitz2}
\ee
From the constraint in (\ref{limitz2}), we will infer the  domain values of $\om$ and $\La$. It is worth mentioning
that in the 3-3-1-1 model, the structure of symmetry breaking which can be divided into  two or three periods depending on scale of VEVs as suggesting in the above two pictures.

\subsection{Two periods EWPT in picture (i)}

In picture (i), we have assumed $\La  \sim \om\gg u\sim v$ meaning that the symmetry breaking or phase transition has two periods.
The first transition is $SU(3)\to SU(2)$ through $\om \sim \La$, which generates masses of the heavy gauge bosons
$X^\pm, Y^\pm, Z_1, Z_2$, Higgs bosons $H_2,H_3, A'_\eta, S'\chi,S_4$, and three exotic quarks. The phase
transition $SU(3)\to SU(2)$ only depends on $\phi_\om\sim \phi_\La$.

When our universe has been expanding and cooling due to $u$ scale, the symmetry breaking $SU(2)\to U(1)$ is turned on,
which generates masses of the SM particles and the last part of masses of $H_2,H_3, X^\pm, Y^\pm$.
Therefore, phase transition $SU(2)\to U(1)$ only depends on $\phi_u\sim \phi_v$.

\subsubsection{Phase transition $SU(3)\to SU(2)$}

This phase transition involves exotic quarks, heavy bosons, but excludes the SM particles. As a consequence,
the effective potential of the EWPT $SU(3)\to U(1)$ is $V_{eff}(\phi_\om)$.

Applying the Coleman-Weinberg's method, the effective potential $V_{eff}(\phi_\om)$ is given as
\be
V_{eff}(\phi_\om) =D_\om(T^2-T_{0\om}^2)\phi_\om^2-E_\om T\phi_\om^3
+\frac{\la_\om(T)}{4}\phi_\om^4 ,\label{t320}
\ee
where
\bea
\la_\om(T)&=&
-\frac{m_{A'_{\eta}}^4\log\left(\frac{m_{A'_{\eta}}^2}{T^2a_b}\right)}{16\pi^2\om^4}-\frac{m_{H_2}^4\log
\left(\frac{m_{H_2}^2}{T^2a_b}\right)}{8\pi^2\om^4}-\frac{m_{H_3}^4\log
\left(\frac{m_{H_3}^2}{T^2 a_b}\right)}{16\pi^2\om^4}-\frac{m_{S'_{\chi}}^4\log
\left(\frac{m_{S'_{\chi}}^2}{T^2a_b}\right)}{16\pi^2\om^4}\crn &
-&\frac{m_{S_4}^4\log\left(\frac{m_{S_4}^2}{T^2a_b}\right)}{16\pi^2\om^4}-\frac{3m_X^4\log
\left(\frac{m_X^2}{T^2a_b}\right)}{8\pi^2\om^4}-\frac{3m_Y^4\log
\left(\frac{m_Y^2}{T^2a_b}\right)}{8\pi^2\om^4}-\frac{3m_{Z_1}^4\log\left(\frac{m_{Z_1}^2}{T^2a_b}\right)}{16\pi^2\om^4}\crn
&-&\frac{3m_{Z_2}^4\log\left(\frac{m_{Z_2}^2}{T^2a_b}\right)}{16\pi^2\om^4}+\frac{3 M_{D_1}^4 \log
\left(\frac{M_{D_1}^2}{T^2a_f}\right)}{4 \pi^2 \om ^4}+\frac{3M_{D_2}^4\log
\left(\frac{M_{D_2}^2}{T^2a_f}\right)}{4\pi^2\om^4}+\frac{3M_U^4\log
\left(\frac{M_U^2}{T^2a_f}\right)}{4\pi^2\om^4}\crn
&+&\frac{m_{A'_{\eta}}^2}{2\om^2}+\frac{m_{H_3}^2}{2\om^2}+\frac{m_{S'_{\chi}}^2}{2\om^2}
+\frac{m_{S_4}^2}{2 \om^2} ,
\label{t3201}\\
E_\om &=&\frac{m_{A'_{\eta }}^3}{12\pi\om^3}+\frac{m_{H_2}^3}{6\pi \om^3}+\frac{m_{H_3}^3}{12\pi\om^3}+\frac{m_{S'_{\chi}}^3}{12\pi\om^3}+\frac{m_{S_4}^3}{12\pi\om^3}+
\frac{m_X^3}{2\pi\om^3}+\frac{m_Y^3}{2\pi\om^3}\\
&+&\frac{m_{Z_1^3}}{4\pi\om^3}
		+\frac{m_{Z_2}^3}{4\pi\om^3} ,
\label{t3202}\\
D_\om &=&\frac{m_{A'_{\eta}}^2}{24\om^2}+\frac{M_{D_1}^2}{4\om^2}+\frac{M_{D_2}^2}{4\om^2}+
\frac{m_{H_2}^2}{12\om^2}+\frac{m_{H_3}^2}{24\om^2}+\frac{m_{S'_{\chi}}^2}{24\om^2}+\frac{m_{S_4}^2}{24 \om^2}+\frac{m_X^2}{4\om^2}+\frac{m_Y^2}{4\om^2}\crn
&+&\frac{m_{Z_1}^2}{8\om^2}
		+\frac{m_{Z_2}^2}{8\om^2}+\frac{M_U^2}{4\om ^2} ,
\label{t3203}\\
F_\om &=&\frac{m_{A'_{\eta}}^4}{32\pi^2\om^2}-\frac{m_{A'_{\eta}}^2}{4}-\frac{3M_{D_1}^4}{8\pi^2\om^2}-
\frac{3M_{D_2}^4}{8\pi^2\om^2}+\frac{m_{H_2}^4}{16\pi^2\om^2}+\frac{m_{H_3}^4}{32\pi^2\om^2}-
\frac{m_{H_3}^2}{4}+\frac{m_{S'_{\chi }}^4}{32 \pi^2\om ^2}-\frac{m_{S'_{\chi }}^2}{4}\crn
&+&\frac{m_{S_4}^4}{32\pi^2\om^2}-\frac{m_{S_4}^2}{4}+\frac{3 m_X^4}{16 \pi^2 \om^2}+\frac{3 m_Y^4}{16 \pi^2\om^2}
+\frac{3m_{Z_1}^4}{32\pi^2 \om^2}
		+\frac{3 m_{Z_2}^4}{32 \pi^2 \om^2}-\frac{3 M_U^4}{8 \pi^2 \om^2} ,
\label{t3204}\eea
and
\be
T_{0\om}^2\equiv -\frac{F_\om}{D_\om} .\label{t3205}
\ee

The minimum conditions are
\be
V_{eff}(0)= \left.\frac{\pa V_{eff}(\phi_\om)}{\pa \phi_\om}\right|_\om=0 ;\quad
	\left.\frac{\pa^2 V_{eff}(\phi_\om)}{\pa \phi_\om^2}\right|_\om=m_{A'_{\eta }}^2+m_{H_3}^2+m_{S'_{\chi }}^2+
m_{S_{4 }}^2 .\label{t3206}
\ee

The values of $V_{eff}(\phi_\om)$ at the two minima become equal at the critical temperature and the phase transition strength are
\bea									
T_{c\om}&=&\frac{T_{0\om}}{\sqrt{1-E_\om^2/D_\om\la_{T_{c\om}}}} ,\crn S_\om
&= &\frac{2E_\om}{\la_{T_{c\om}}} .
\nn \eea

From Eqs. (\ref{kluongznhe}, \ref{kluongz1nhe}, \ref{kluongz2nhe}), with the limit of $m_{Z_2}$ given in Eq.(\ref{limitz2}),
it follows:  $5.856 \text{ TeV}\le \om\sim \La \le 6.654$ TeV.

\begin{figure}[!h]
		\includegraphics[scale=1.5]{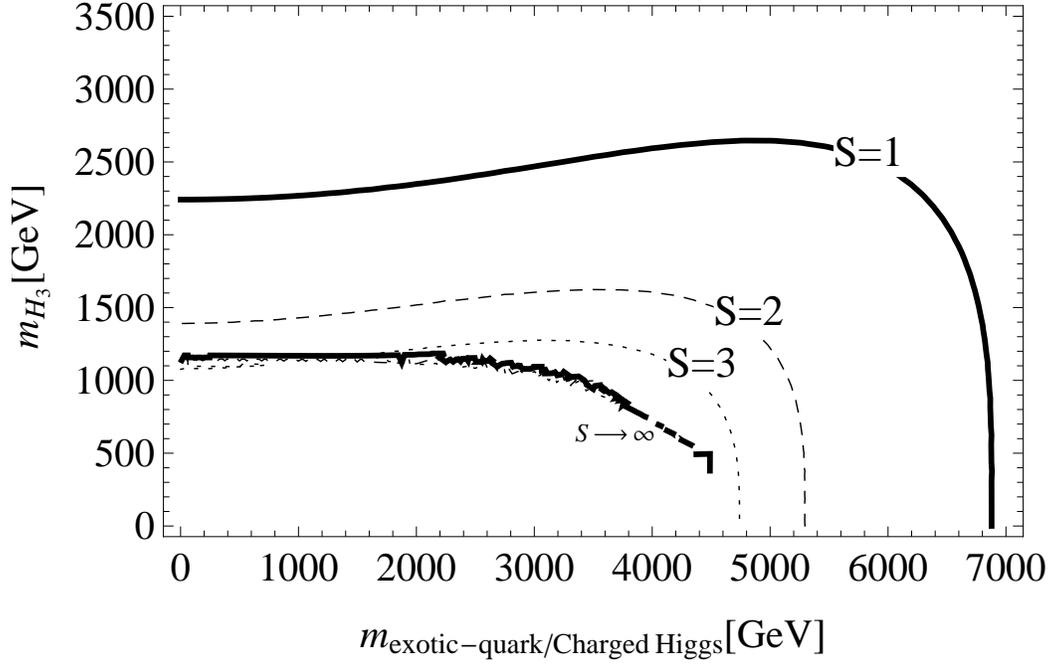}
		\caption{The mass area corresponds to $S_{\om}>1$}
	\label{h311}
\end{figure}

\begin{figure}[!ht]
	\includegraphics[scale=1.5]{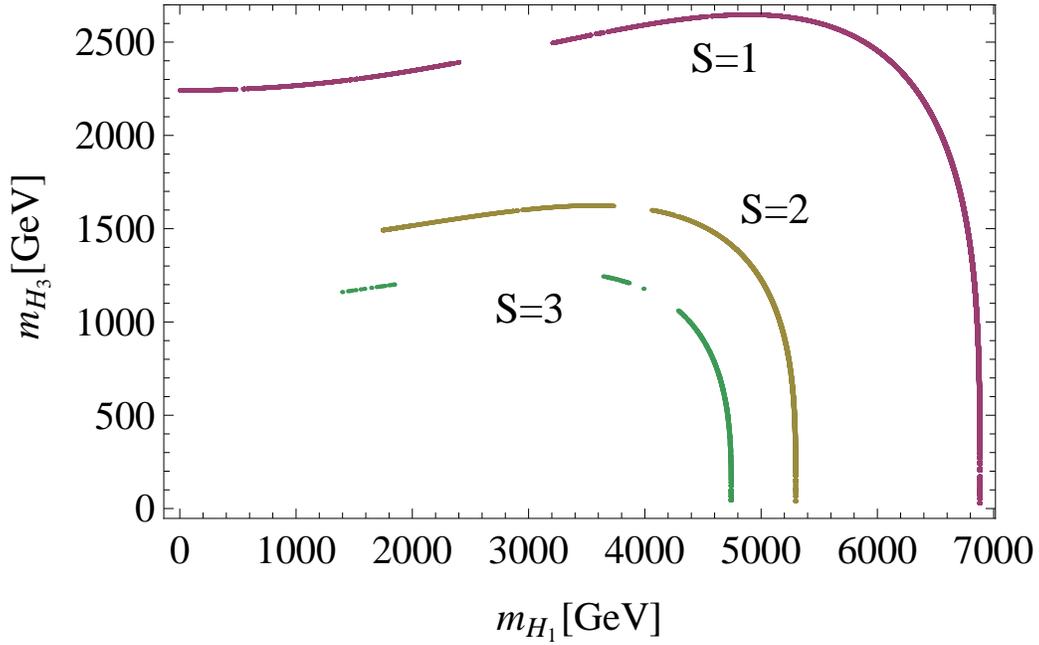}
	\caption{The mass area corresponds to $S_{\om}>1$ with real $T_C$ condition. The gaps on the lines ($S=1, 2, 3$)
correspond to values  making $T_C$ to be complex.}
	\label{h312}
\end{figure}

In this work, we assume $\om=6$ TeV, so that $m_{Z_1}=8.304\, \text{ TeV}$ and $m_{Z_2}=2.254\,\text{ TeV}$.
The problem here is that there are nine variables: the masses of $U, D_1,D_2,H_2,H_3$ and $A'_\eta, S'_\chi,S_4,Z_1$.
However, for simplicity, we assume $m_U=m_{D_1}=m_{D_2}=m_{H_2}\equiv O$, $m_{A'_\eta}=m_{S'_\chi}=m_{H_3}=m_{S_4}\equiv P$.
Consequently, the critical temperature and the phase transition strength are the function of $O$ and $P$; therefore we can
rewrite the phase transition strength as follows
\be
S_\om=\frac{2 E_\om}{\la_{T_{c\om}}}\equiv S_\om(O,P,S_\om) .\label{str11}
\ee
In Figs. \ref{h311} and \ref{h312}, we have plotted the relation between masses of the charged particles $O$ and neutral particles $P$
with some values of the phase transition strength at $\om=6$ TeV.

The mass region of particles is the largest at $S_\om=1$, the mass region of charged particles and neutral particles are
\[
\left\{\begin{split}&0\leq m_{Exotic Quark/Charged Higgs boson}\leq 7000 \text{ \gev},\\& 0\leq m_{H_3}\leq 2600 \text{ \gev} .\end{split}\right.
\]
From Eq. (\ref{str11}) it follows that the maximum of $S_\om$ is around  70.

\subsubsection{Phase transition $SU(2)\to U(1)$}

In this period, the symmetry breaking scale equals to $u=246/\sqrt{2}$ and
 the masses of the SM particles and apart of masses of $ X,Y,H_1, H_2, H_3,A_\chi, S_\eta$ are generated.

There are six variables corresponding to the masses of bosons $H_1,H_2,A_\chi, A_\eta, H_3,S_\rho$. For simplicity,
we assume: $m_{H_1}=m_{H_2}\equiv K$, $m_{A_\chi}=m_{S_\eta}=m_{H_3}\equiv L$, and $m_{S_\rho}=125$ GeV.

\begin{figure}[!ht]
		\includegraphics[scale=1.5]{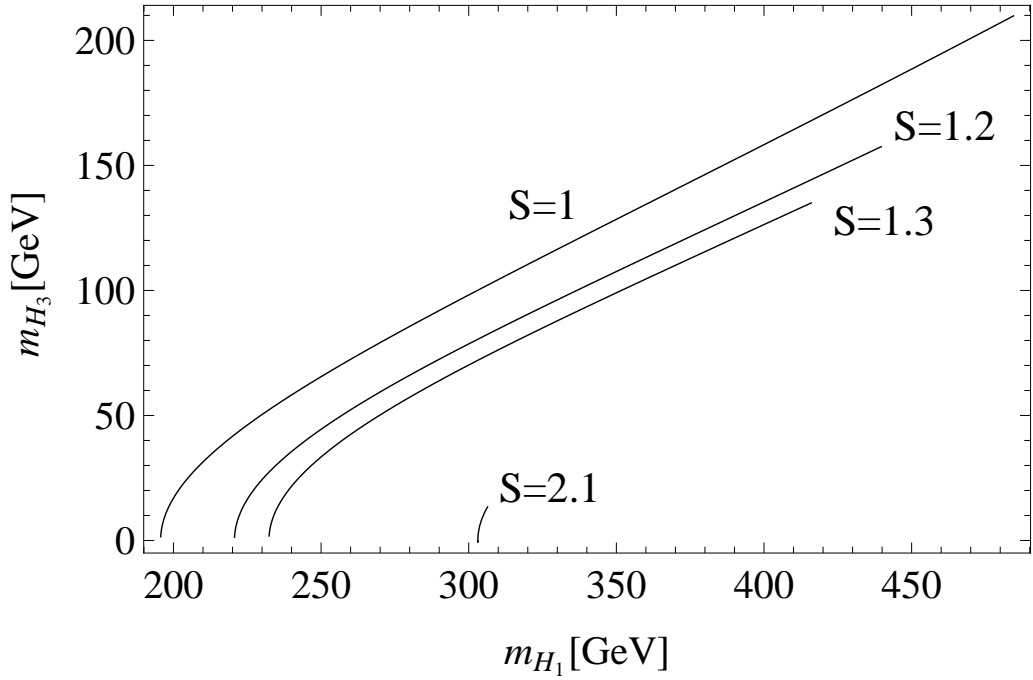}
		\caption{The strength EWPT $S=\dfrac{2E_u}{\la_{T_c}}$ with $T_c$ must be real.}
	\label{h33c}
\end{figure}

The effective potential of EWPT $SU(2)\to U(1)$ is given as
\[
V_{eff}(\phi_u)=\frac{\la_u(T)}{4}\phi_u^4-E_uT\phi_u^3+D_uT^2\phi_u^2+F_u\phi_u^2 .
\]
The minimum conditions are
\be
V_{eff}(0)= \left.\frac{\pa V_{eff}(\phi_u)}{\pa \phi_u}\right|_u=0;\quad \left.\frac{\pa^2 V_{eff}(\phi_u)}{\pa \phi_u^2}\right|_u
=m_{A_{\chi }}^2+m_{H_3}^2+m_{S_{\eta }}^2+m_{S_{\rho }}^2 ,\label{chuanhoa1}
\ee
where
\bea
D_u &=&\frac{m_{A_{\chi}}^2}{24u^2}+\frac{m_{H_1}^2}{12u^2}+\frac{m_{H_2}^2}{12 u^2}+\frac{m_{H_3}^2}{24u^2}+\frac{m_{S_{\eta }}^2}{24u^2}+\frac{m_{S_{\rho}}^2}{24u^2}+\frac{m_W^2}{4u^2}+\frac{m_X^2}{4u^2}+\frac{m_Y^2}{4u^2}\crn
&+&\frac{m_Z^2}{8 u^2}+\frac{M_t^2}{4u^2},\crn
F_u & =&\frac{m_{A_{\chi}}^4}{32\pi^2u^2}-\frac{m_{A_{\chi}}^2}{4}+\frac{m_{H_1}^4}{16\pi^2u^2}+\frac{m_{H_2}^4}{16\pi^2u^2}
+\frac{m_{H_3}^4}{32\pi^2u^2}-\frac{m_{H_3}^2}{4}-\frac{m_{S_{\eta}}^2}{4}-\frac{m_{S_{\rho}}^2}{4}
\crn
&+&\frac{m_{S_{\eta}}^4}{32 \pi^2u^2}+\frac{m_{S_{\rho }}^4}{32\pi^2u^2}+\frac{3 m_W^4}{16 \pi^2u^2}
+\frac{3 m_X^4}{16 \pi^2u^2}+\frac{3 m_Y^4}{16 \pi^2u^2}+\frac{3m_Z^4}{32 \pi^2u^2}-\frac{3 M_t^4}{8 \pi^2 u^2},\crn
E_u &=&\frac{m_{A_{\chi }}^3}{12 \pi u^3}+\frac{m_{H_1}^3}{6 \pi u^3}+\frac{m_{H_2}^3}{6 \pi u^3}
+\frac{m_{H_3}^3}{12 \pi u^3}+\frac{m_{S_{\eta }}^3}{12 \pi u^3}+\frac{m_{S_{\rho}}^3}{12 \pi u^3}
+\frac{m_W^3}{2 \pi u^3}+\frac{m_X^3}{2\pi u^3}\crn
&+&\frac{m_Y^3}{2 \pi u^3}+\frac{m_Z^3}{4 \pi u^3},\crn
\la_u(T)&=&-\frac{m_{A_{\chi }}^4 \log \left(\frac{m_{A_{\chi }}^2}{T^2a_b}\right)}{16 \pi ^2 u^4}-
\frac{m_{H_1}^4 \log\left(\frac{m_{H_1}^2}{T^2 a_b}\right)}{8 \pi^2u^4}-\frac{m_{H_2}^4 \log \left(\frac{m_{H_2}^2}{T^2a_b}\right)}{8 \pi ^2 u^4}-\frac{m_{H_3}^4 \log\left(\frac{m_{H_3}^2}{T^2 a_b}\right)}{16 \pi^2u^4}\crn
&- &\frac{m_{S_{\eta }}^4 \log \left(\frac{m_{S_{\eta}}^2}{T^2 a_b}\right)}{16 \pi ^2 u^4}-
\frac{m_{S_{\rho }}^4\log \left(\frac{m_{S_{\rho }}^2}{T^2 a_b}\right)}{16 \pi^2 u^4}-
\frac{3 m_W^4 \log \left(\frac{m_W^2}{T^2a_b}\right)}{8 \pi ^2 u^4}-\frac{3 m_X^4 \log\left(\frac{m_X^2}{T^2 a_b}\right)}{8 \pi ^2 u^4}\crn
&- &\frac{3m_Y^4 \log \left(\frac{m_Y^2}{T^2 a_b}\right)}{8 \pi^2u^4}-\frac{3 m_Z^4
 \log \left(\frac{m_Z^2}{T^2a_b}\right)}{16 \pi ^2 u^4}+\frac{3 M_t^4 \log\left(\frac{M_t^2}{T^2 a_f}\right)}{4 \pi^2u^4}\crn
&+&\frac{m_{A_{\chi }}^2}{2 u^2}+\frac{m_{H_3}^2}{2u^2}+\frac{m_{S_{\eta }}^2}{2 u^2}+\frac{m_{S_{\rho }}^2}{2u^2} .
\nn
\eea
The critical temperature and the phase transition strength are given by
\bea
T_c&=&\frac{T_0}{\sqrt{1-\frac{E^2}{D\la_{T_c}}}} ,\crn
 S &= &\frac{2E}{\la_{T_c}} .
 \label{str}\eea
Like the phase transition $SU(3)\to SU(2)$, in Fig. \ref{h33c} we have plotted the relation
between masses of the charged particles $K$
 and neutral particles $L$ with some values of the phase transition strength.

\begin{figure}[!ht]	
		\includegraphics[scale=1.5]{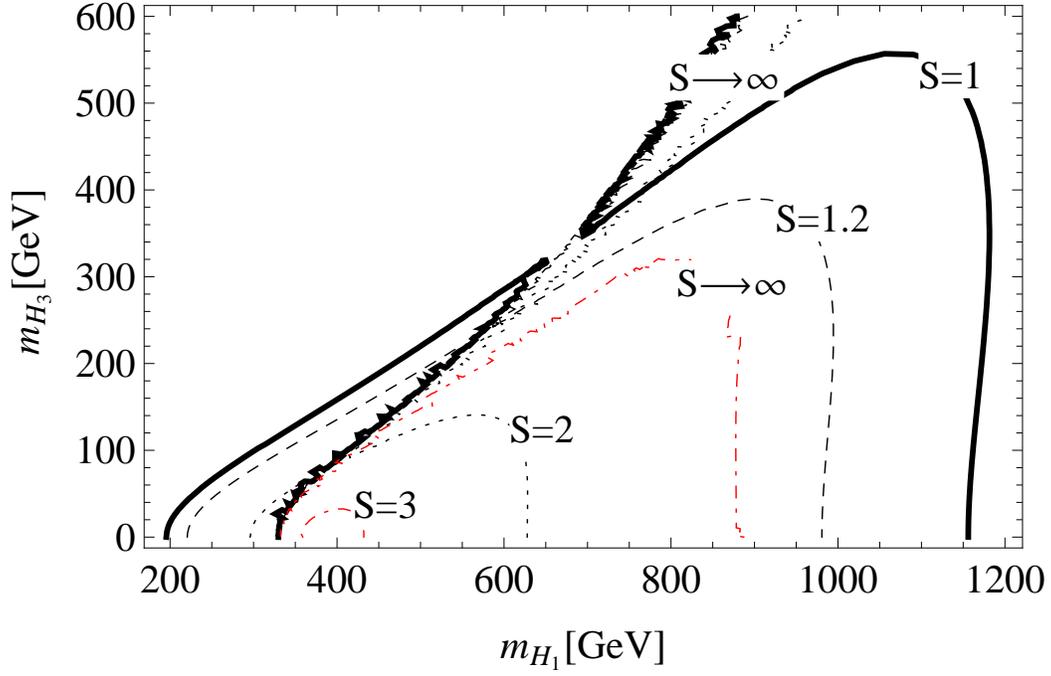}
		\caption{The strength $S=\dfrac{2E_u}{\la_{T_c}}$.}
	\label{h33a}
\end{figure}

However,  we can fit the mass of heavy particle one again when considering the
 condition of $T_c$ to be real, so that Fig. \ref{h33a} is
redrew to Fig.\ref{h33c} and the maximum of strength is reduced from 3 to 2.12.

\begin{table}[!ht]
	\caption{Mass limits of particles with $T_C>0$.}
	\scalebox{1.2}{
		\begin{tabular}{ l|c|c}
			\hline
			Strength $S$ &$K[\gev]$&$L[\gev]$\\\hline
			1.0-2.12&$195\leq K\leq 484.5$&$0\leq L\leq 209.8$\\
			\hline
		\end{tabular}}\label{fit}
\end{table}
	
The mass region of neutral and charged particles given in Table (\ref{fit}) leads the maximum phase
transition strength which must be $2.12$.
This is larger than 1 but  the EWPT is not strong.
	
%\begin{figure}[!ht]
%		\includegraphics[scale=1]{cp23.eps}		
%		\caption{The dependence of the effective potential $V_{eff}(u)$ on the temperature with $m_{H_3}=118.6 \gev$, $m_{H_1}=333.6 \,
% \gev$, $T_c=125.88 \, \gev$, and $S=1$.}
%		\label{h35}
%\end{figure}
	
\subsection{Three period EWPT in picture $(ii)$}

In picture (ii), $m_{Z_2}^2\simeq \fr{g^2c_W^2 \om^2}{(3 - 4 s_W^2)}$ with the limit of $m_{Z_2}$ given  in Eq. (\ref{limitz2}),
we obtain $5.53 \text{ TeV}\le \om \le 6.3$ TeV. Therefore, we also assume $\om=6$ TeV in this picture.

Because $\La \gg \om=6$ TeV and $\om\gg u\sim v$, therefore there are three periods. The first process is
$SU(3)_L\otimes U(1)_X \otimes U(1)_N \longrightarrow SU(3)_L\otimes U(1)_X$. The second one is $SU(3)_L\otimes
U(1)_X\longrightarrow SU(2)_L\otimes U(1)_X$. The third process is $SU(2)_L\longrightarrow U(1)_Q$. The third process is
 like $SU(2)\longrightarrow U(1)$ in the picture (i).

The first process is a transition of the symmetry breaking of $U(1)_N$ group. It generates mass for $Z_1$ through $\La$ or Higgs
boson $S_4$. The third process is like the $SU(2)\longrightarrow U(1)$ EWPT in picture (i). The second process is like
the $SU(3)\longrightarrow SU(2)$ in picture (i) but it does not involve $Z_1$ and $S_4$.

The second process has the effective potential is like Eq. (\ref{t320}).
In addition,  parameters and the minimum conditions are
 like Eqs. (\ref{t3201},\ref{t3202},\ref{t3203},\ref{t3204},\ref{t3205},\ref{t3206}) without $Z_1$ and $S_4$.

\begin{figure}[!ht]	
		\includegraphics[scale=1.5]{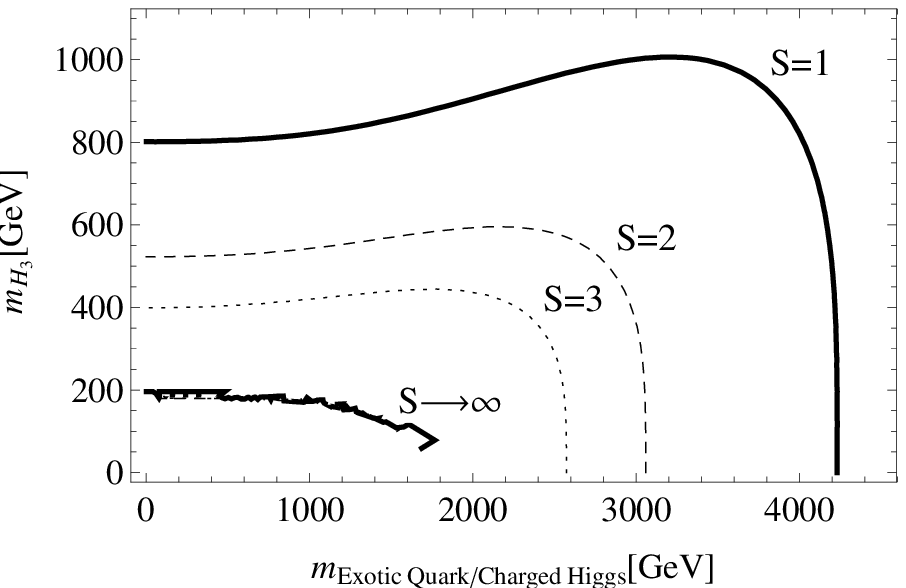}
		\caption{The strength EWPT $S_{\om}=\dfrac{2E_\om}{\la_{T_c}}$ with $\om=6$ TeV.}
	\label{h33}
\end{figure}

In our numbering process, when we import real $T_C$, the mass region of charged and neutral particles are
\bea\left\{\begin{split}&0\leq m_{Exotic quark/Charged Higgs boson}\leq 4000 \text{ \gev},\\& 0\leq m_{H_3}\leq 1000 \text{ \gev} .\end{split}\right.\nn \eea
The mass region of charged bosons is narrower than that in  Fig. \ref{h33}. From Eq. (\ref{str11}), the maximum of $S$ has been estimated to be around 100.

\section{The role of neutral fermions in EWPT}\label{sec5}

The masses of $N_R$ can be generated by the scalar content by itself via an effective operator  invariant under the 3-3-1-1 symmetry and $W$-parity \cite{3311}:
\be
\fr{\la_{ab}}{M}\bar{\psi}^C_{aL} \psi_{bL}(\chi^\dag \chi ) .
\label{M}
\ee
The mass scale of $N_R$ is unknown, but  it can be taken in TeV scale.  However,
the analysis   of the
 scattering of $N_R$
with distributions of $X, Y, Z_2$ bosons
 given in \cite{3311} leads to a consequence that the mass of $N_R$ is  equal or less than that of the $Z_2$ boson  as follows
\be m_{N_R}=\frac{m^2_{Z_2}}{2.557\text{ TeV}}\le m_{Z_2} .\label{limitnr}\ee

In the $SU(3)\longrightarrow SU(2)$, if we add the contribution of neutral fermions, then the maximum of $S$
would decrease.
However,  the neutral fermions do not make lose the first-order EWPT
 as  shown in   Table \ref{smax}

\begin{table}[!ht]
	\caption{Values of the maximum of EWPT strength with $\om=6$ TeV.}
	\scalebox{1.2}{
		\begin{tabular}{ l|l|c|c|c|c}
			\hline
			 Period&Picture &$m_{Z_2}$[TeV]&$m_{N-R}$[TeV]&$S_{Max}$ without $N_R$&$S_{Max}$
			 with$N_R$\\\hline
			$SU(3)\longrightarrow SU(2)$&$(i)$&2.386&2.227&70&50\\
			$SU(3)\longrightarrow SU(2)$&$(ii)$&2.254&1.986&100&30\\
			\hline
		\end{tabular}}\label{smax}
	\end{table}
	
In Table \ref{smax}, we have only estimated the maximum strengths and showed that these maximum values are significantly reduced. However,
 it is very difficult to  calculate these values accurately because of the existence of
  many parameters (the masses  of heavy particles);
 and these values can change slightly (but not too much) with different approximations.
Looking at  the Table \ref{smax},  the following  remarks  are in order:
\ben
\item
 In case of the neutral fermion absence. In the picture (i), if $Z_1$ boson  is involved in the $SU(3)\longrightarrow SU(2)$ EWPT; the contribution of $Z_1$ makes increasing $E$ and $\la$, but $\la$ increases stronger than $E$. The strength $S=\frac{2E}{\la_{T_c}}$ gets the value equals 70.
  For the picture $(ii)$,  the mentioned  value equals 100.
\item
In case of the neutral fermion existence.
When the  neutral fermions are involved in both  pictures, $S_{max}$ in picture $(ii)$ decreases faster
than $S_{max}$ in picture $(i)$.  The strength gets values equal to 50 and 30 for the picture $(i)$ and $(ii)$, respectively.
\een

If the neutral fermions follow the Fermi-Dirac distribution (i.e., they act as a real fermion but without lepton number), they increase the value
of the $\la$ and $D$ parameters. Thus, they reduce the value of strength EWPT $S$, because $S=\frac{E}{2\la_{T_c}}$
and $E$ do not depend on the neutral fermions.

This suggests that DM candidates
 are neutral fermions (or fermions in general) which  reduce the maximum value of the EWPT strength.

However, the EWPT process depends on bosons and fermions. Boson gives a positive contribution (obey the Bose-Einstein distribution)
 but fermion gives a negative contribution (obey the Fermi-Dirac distribution). In order to have the first order transition, the symmetry
 breaking process must generate mass for more boson than fermion.

In addition, in this model, the neutral fermion mass is generated from an effective operator. This operator which demonstrates an interaction
 between neutral fermions and two Higgs fields. The above neutral fermion is very different from usual fermions.
The $M$ parameter in (\ref{M}) has an energy dimension,
 and it may be an un-known dark-interaction. Thus, the neutral fermions only are
 effective fermions, according to the Fermi-Dirac distribution, but their  statistical nature needs to be further analyzed with other data.

\section{Conclusion and outlooks}\label{sec6}

In this paper, we have considered the EWPT  in the  3-3-1-1 model where the SSB can be separated into two or three scales.
 Hence, in the model under consideration, the EWPT  consists of  two  pictures.
The first picture containing  two periods of  EWPT,  has a transition $SU(3) \rightarrow  SU(2)$ at 6 TeV scale and another is
$SU(2) \rightarrow U(1)$ transition which is the like-standard model EWPT.
The second picture is an EWPT structure containing  three periods, in  which  two first periods are similar to those of the first picture and another one is the symmetry breaking process of $U(1)_N$ subgroup. The EWPT is
  the first order phase transition if new bosons with mass within range of some TeVs  are triggers for the purpose.
 The maximum strength of the $SU(2) \rightarrow U(1)$ phase transition is equal to 2.12 so the EWPT is not strong.

We have focused on the  neutral fermions without lepton number being candidates for DM and obey the Fermi-Dirac distribution, and have shown that the mentioned fermions can be a negative trigger for EWPT.
       Furthermore, in order to be the strong first-order EWPT at TeV scale, the symmetry breaking processes must produce more bosons than fermions or the mass of bosons must be much larger than that of fermions.

It is known  that the mass of Goldstone boson is very small \cite{pdg} and the physical quantities are gauge independent so the critical
 temperature and the strength is gauge independent \cite{1101.4665}.
Consequently, the survey of effective potential in Landau gauge is also sufficient, or
 other word speaking, it is just consider in determined gauge.
 Thus,  it is a reason why  the Landau gauge is used in this work.  In  this paper, the structure of EWPT in the 3-3-1-1 model with the effective potential at finite temperature has been  drawn at the 1-loop level; and this potential has two or three phases.

We have analyzed the processes which generate the masses for all gauge bosons inside the covariant derivatives. After diagonalization, the masses of gauge bosons do not have   mixing among VEVs. Therefore, the EWPT stages are independent of each other  \cite{phonglongvanE}.

To avoid higher (six) order Higgs self-interaction in the effective potential, the $f$ parameter associated with triple scalar atisymmetric coupling  is ignored. Thus calculating the corrections with $f$ can reveal many new physical phenomena. In addition, from the phase transitions, we can get some bounds on the Higgs self-couplings.

In conclusion, the model has many bosons which will be good triggers for first-order EWPT.  The situation is that as  less heavy fermion as the result will be  better. However, strength of EWPT can be reduced by many bosons (such as $Z, Z_1, Z_2$ in the 3-3-1-1 model).

The new scalar particles playing  a role in generation mass  for exotic particles, increase the value of EWPT strength. Because these scalar fields follow the Bose-Einstein distribution, so that they contribute positively to the effective potential. With  the help of such particles, the strength of phase transition will be strong. As mentioned above, their masses depend just on  one VEV, so they only participate in one phase transition. Moreover, among with the neutral fermion, they may be  candidates  for DM. From the point of view of the early universe, the above particles can be an inflaton or some product of the inflaton decay.

Although we only work on the 3-3-1-1 model, but this manipulation can still apply to other models with multi-period EWPT.  We find that the results about bosons in Ref.~\cite{Furey:2018drh} or new models (with $SU(5)$ or $SU(6)$ groups) in Ref.~\cite{Deppisch:2016jzl}, can be a benchmark or may contain new material for the problem considered here-triggers for EWPT.

The heavy neutrinos or quarks mixing, in Ref.~\cite{Fonseca:2016tbn}, are an interesting issue, and  they may be the source for CP violations.  In order to analysis in details baryogenesis, our next works will  be consideration of CP violations and correction of neutral fermion-dark matter.

The model under consideration  is an extension of the SM symmetry group,  so it  is renormalizable and there are no Landau poles when choosing the appropriate parameters. Inflation and kinetic mixing effect via $\rho$ parameter have been performed in Ref [24, 25, 26]. We will  also perform one UV completion of this model without the  $f$ term in the Higgs potential.

It is interesting to note that the bound ($w > 3.2$ TeV) obtained  here from the EWPT  is  consistent  with those followed from the oblique corrections in Ref.~\cite{3311Dsi}.

The largest cutoff of this model is $\La$, may not be option. In addition, energy scale of the model goes from high to low ($\La \longrightarrow \om \longrightarrow u \sim v$) so that the model has two cutoff scales which larger than $246$ GeV. This is a common thing of all beyond SM.
	
We see that this model is correct for the 246 GeV energy scale;  the model has materials for the first order EWPT. However, this does not confirm that the model is correct at arbitrary energy level which requires further study/experimentation.
	
We also recognize that phases occur at different energy scales. The UV completion from the low to high scale, has been not clearly linked. In order to construct  model, we need to consider EWPT, because the EWPT will  make the appearance of UV completion. Therefore, in the next work with 3-3-1-1 model revisited, we will correct the model in combining with the UV completion for the Higgs potential as in Ref.~\cite{uv}.

\section*{Acknowledgment}
This research is funded by Vietnam  National Foundation for Science and Technology Development (NAFOSTED) under grant number 103.01-2017.356.
\\[0.3cm]

\end{document}